\documentclass{lmcsTLengl}
\usepackage[latin1]{inputenc}
\usepackage{longtable}
\title[A Decision Procedure for Herbrand Formulae]{A Decision Procedure for Herbrand Formulae without Skolemization}
\author[Lampert]{Timm Lampert}
\address{Humboldt University Berlin\\
	Unter den Linden 6\\
	D-10099 Berlin}
\email{lampertt@staff.hu-berlin.de}
\begin{document}

\begin{abstract}
This paper describes a decision procedure for disjunctions of conjunctions of anti-prenex normal forms of pure first-order logic (FOLDNFs) that do not contain $\vee$ within the scope of quantifiers. The disjuncts of these FOLDNFs are equivalent to prenex normal forms whose quantifier-free parts are conjunctions of atomic and negated atomic formulae (= Herbrand formulae). In contrast to the usual algorithms for Herbrand formulae, neither skolemization nor unification algorithms with function symbols are applied. Instead, a procedure is described that rests on nothing but equivalence transformations within pure first-order logic (FOL). This procedure involves the application of a calculus for negation normal forms (the NNF-calculus) with
$A \dashv\vdash A \wedge A$ (= $\wedge$I) as the sole rule that increases the complexity of given FOLDNFs.
The described algorithm illustrates how, in the case of Herbrand formulae,
decision problems can be solved through a systematic search for proofs that
reduce the number of applications of the rule $\wedge$I to a minimum in the NNF-calculus. In the case of Herbrand formulae, it is even possible to entirely abstain from applying $\wedge$I.
Finally, it is shown how the described procedure can be used within an optimized general search for
proofs of contradiction and what kind of questions arise for a $\wedge$I-minimal proof strategy in the case of a general search for
proofs of contradiction.
\end{abstract}

\maketitle


\section{Introduction}

\begin{defi}
A \emph{Herbrand formula} is a formula in prenex normal form whose matrix is a conjunction of atomic or negated atomic formulae.
\end{defi}

\cite{Boerger}, section 8.2.2, describe a decision procedure for the refutability of Herbrand formulae that reduces the decision problem for such formulae to the unification problem, which is decidable; cf. \cite{Boerger}, Theorem 8.2.1.
This reduction presumes skolemization. Resolution decision procedures that are based on the resolution calculus use skolemization and unification algorithms to decide further decidable classes of first-order logic, cf. \cite{Fermueller1} and \cite{Fermueller2}. 

As an alternative to this approach, I will describe in the following how the refutability of Herbrand formulae can be decided without skolemization. 
I will refer to nothing but equivalence transformations of negation normal forms (NNFs) of pure first-order logic (FOL), without names, free variables, function symbols or identity. The problem for the refutability of Herbrand formulae will also be reduced to a problem of unifying literals. However, unifying literals through a logically valid proof procedure within the NNF-calculus amounts to answering the question of to what extent universally quantified variables can be replaced with existentially quantified variables in NNFs. 
The leading idea of the decision procedure is to reduce the application of the rule $\wedge$I in proofs within the NNF-calculus to a maximal extent. In case of the Herbrand formulae, it is even possible to entirely abstrain from applying $\wedge$I. The specification of the decision procedure for the rather simple case of Herbrand formulae described in this paper shall serve as a starting point for further investigations that consider more complicated cases. The motivation of this paper is the conviction that the strategy to reduce the application of $\wedge$I in the search of a proof within the NNF-calculus to a finite number
opens up new perspectives in the search of decision procedures for FOL.

I begin from FOL formulae expressed as NNFs.
\begin{defi}
\emph{Negation normal forms} (NNFs) are formulae that contain only $\neg$, $\wedge$ and $\vee$ as sentential connectives, with $\neg$ occurring only directly in front of atomic expressions.
 \end{defi}
Any formula of FOL can be converted into an NNF through equivalence transformations.\footnote{Cf., e.g., \cite{Nonnengart}, section 3.2.}

%\begin{algo}\hfill
%\begin{item}
%\item Bilde die Skolem Form von $\phi$.
%\item Ziehe alle Allquantoren nach innen.
%\item Indiziere maximal.
%\item Bilde die pränexe Normalform.
%\item Entscheide, ob ein konnektiertes Literalpaar unifizierbar ist.
%\end{algo}

Herbrand formulae are NNFs in prenex normal form. There is no need to presume prenex normal forms as long as $\vee$ does not occur in the NNFs of interest.
\begin{defi}
\emph{$\wedge$-NNFs} are NNFs without $\vee$.
\end{defi}

Any disjunction is contradictory iff each disjunct is contradictory. Therefore, a disjunction is decidable if each disjunct is decidable. Consequently, the greatest profit in applying a decision procedure for $\wedge$-NNFs arises from driving $\vee$ outward as far as possible by generating FOLDNFs, i.e., disjunctions of conjunctions of anti-prenex normal forms, which are expressed as NNFs. For a convenient algorithm for converting any FOL formula into a FOLDNF of minimal length 1, cf. \cite{Lampert}. The procedure described within this paper enables a decision regarding whether a FOLDNF $A$ is contradictory if each disjunct of $A$ is a $\wedge$-NNF.

The decision problem for Herbrand formulae or $\wedge$-NNFs can \label{redphitopsi} be reduced to a decision problem for $\wedge$-NNFs $\psi$ with only two literals.\footnote{
I refer to \emph{two occurrences} of literals and not to an arbitrary number of occurrences of \emph{two types} of literals.}
\begin{defi}
The \emph{subformulae} $\psi$ of a $\wedge$-NNF $\phi$ are formulae that are generated by deleting all but two literals, L1 and L2, in $\phi$ and retaining only those quantifiers that bind variables in L1 and L2.
\end{defi}

\begin{thm}\label{subform}
A $\wedge$-NNF $\phi$ is contradictory iff at least one of its subformulae $\psi$ is contradictory.
\end{thm}

\proof
Any proof of the refutability of a $\wedge$-NNF $\phi$ in the tree calculus (tableaux) can be reduced to a
proof of the refutability of $\psi$;
all that must be done is to eliminate quantifiers in the same manner as in the proof of the refutability of $\psi$.
\qed

I will describe in sections \ref{minscope} to \ref{bkrit} how one can decide whether a subformula $\psi$ is contradictory by considering whether the pair $\{L1,L2\}$ is unifiable.
In the preceding sections \ref{unifikation} to \ref{undmin}, I will establish general presuppositions for proofs of contradiction for NNFs.
In sections \ref{optwid} to \ref{subkpairs}, I will explain how to make use of the decision regarding the possibility of unifying $\{L1,L2\}$ to optimize the search for general
proofs of contradiction.
Finally, in section \ref{ausblick}, I will consider what questions arise when generalizing the strategy of the described decision procedure.

\section{Unification}\label{unifikation}

Proofs of contradiction depend on the unification of pairs of connected literals. This section explains what that means.

I will use subscripted $x$ variables (`$x$ variables' for short) to represent variables in NNFs that are bound by universal quantifiers and subscripted $y$ variables (`$y$ variables' for short) to represent those bound by existential quantifiers.
Furthermore, I assume that the NNFs are rectified.
\begin{defi}
\emph{Rectified} NNFs are NNFs in which any two quantifiers bind different variables.
\end{defi}
Any NNF can be transformed into a rectified NNF with universally quantified $x$ variables and existentially quantified $y$ variables. To do so, the variables must be renamed by means of SUB1 and SUB2; cf. table \ref{subg}.

\begin{longtable}{rclll}
$\exists \mu A(\mu)$ & $\dashv\vdash$ & $\exists \nu A(\nu/\mu)$ & Restriction: $\nu$ does not occur in $A(\mu)$  & SUB1\\
$\forall \mu A(\mu)$ & $\dashv\vdash$ & $\forall \nu A(\nu/\mu)$ & Restriction: $\nu$ does not occur in $A(\mu)$ & SUB2\\
\caption{Laws of Substitution}\label{subg}
\end{longtable}

The question of unifiability is whether $x$ variables can be replaced with $y$ variables in a rectified NNF such that identical $y$ variables occupy identical positions\footnote{`Identical positions in two literals L1 and L2' means `the same position in the propositional functions involved'. Thus, I ignore the negation sign that might occur in a literal. In $\{Fx_{1}y_{2},\neg Fy_{1}y_{2}\}$, for example, $x_{1}$ and $y_{1}$ are in identical positions, as are the two occurrences of $y_{2}$.} in connected literals.
A pair of connected literals of a rectified NNF is defined as follows:

\begin{defi}\label{konnektiert}
Two literals L1 and L2 of a rectified NNF with universally quantified $x$ variables and existentially quantified $y$ variables constitute a \emph{pair of connected literals} iff
\begin{enumerate}
\item L1 is not negated, whereas L2 is negated.
\item L1 and L2 contain the same predicate letter $\varphi$.
\item The same number of positions succeed $\varphi$.
\item If position $n$ in L1 is occupied by $y_{i}$, then position $n$ in L2 is not occupied by $y_{j}$ for $i \neq j$.
\end{enumerate}
\end{defi}

Manifestly, these conditions are necessary for L1 and L2 to constitute a unifiable pair of literals that might contribute to
a proof of contradiction.
A rectified NNF that does not contain any pair of connected literals cannot be contradictory.
Thus, the decision problem for subformulae $\psi$ is reduced to the corresponding problem for those subformulae $\psi$ that contain a pair of connected literals L1 and L2.
In sections \ref{minscope} to \ref{bkrit}, I will consider only subformulae $\psi$ of this kind.

I adopt the following narrow understanding of unification.
\begin{defi} \label{defunification} The \emph{unification} of a pair of connected literals in an NNF is the logically valid procedure of replacing $x$ variables such that identical positions in a pair of connected literals are occupied by identical $y$ variables. A pair of connected literals is \emph{unified} if all of its positions are occupied by identical $y$ variables.
A pair of connected literals is \emph{unifiable} if it is possible to unify it through unification.
\end{defi}
It is presumed that during unification, \emph{all} universally quantified $x$ variables are ultimately eliminated, even in cases such as $\{Fx,\neg Fx\}$.

Unification in the sense of Definition \ref{defunification} is independent of skolemization. In the context of rectified NNFs, nothing matters for unification but the replacement of $x$ variables with $y$ variables. Unification must obey logical laws. The problem of unification concerns the replacement of $x$ variables with $y$ variables by means of universal quantifier elimination within a correct and complete calculus for NNFs. Before explaining a procedure for solving this problem, I will first establish the calculus that I will adopt for %proofs by contradiction
proofs of contradiction for NNFs. I refer to a calculus that applies not only to $\wedge$-NNFs but to NNFs in general because it is desired that it should be possible to extend the considered proof strategies to NNFs in general.

\section{NNF-Calculus}\label{nnfkalkuel}

This section establishes a correct and complete calculus for NNFs that I call the `NNF-calculus'.
As in the tree calculus, proofs in the NNF-calculus are
proofs of contradiction: what is to be proven is not that a formula is a theorem but that a formula is a contradiction.
Unlike in the tree calculus, however, neither free variables nor names are used.
Instead, any proof operates within the realm of \emph{pure} FOL.
Also, in contrast to the tree calculus, the NNFs are not decomposed within proofs in the NNF-calculus.
Before universal quantifiers are eliminated for unification, equivalence transformations are applied.
Universal quantifier elimination is the last step in a proof of contradiction for the sake of unification in the case that a formula is contradictory;
cf. section \ref{beweisschritte}.

The rule for universal quantifier elimination is as follows:
\begin{longtable}{rcll}
$\exists \mu \ldots \forall \nu A(\mu,\nu)$ & $\vdash$ & $\exists \mu A(\mu,\nu/\mu)$ & $\forall$E\\
\caption{Universal Quantifier Elimination}
\end{longtable}

If $\forall \nu$ is within the scope of $\exists \mu$, then the universal quantifier $\forall \nu$ is eliminated, and all occurrences of $\nu$ in the scope of $\forall \nu$ are replaced with $\mu$.
It is also allowed to replace $\nu$ with $\mu$ when $\exists \mu$ is a \emph{new} existential quantifier preceding the resulting formula. I subsume this case under $\forall$E. I arbitrarily choose the variable $y_{0}$ as a new $y$ variable and require that $y_{0}$ does not occur in the expression to the left of $\vdash$ in $\forall$E.
\label{y0}

In addition to $\forall$E, the following rule is crucial for %proofs by contradiction
proofs of contradiction in the NNF-calculus:
 \begin{longtable}{rcll}
$A$ & $\dashv\vdash$ & $A \wedge A$ & $\wedge$I\\
 \caption{Conjunct Multiplication}
 \end{longtable}
In proofs of contradiction in the NNF-calculus,
$\wedge$I is used to multiply universally quantified expressions to the effect that $x$ variables of different conjuncts are replaced with different $y$ variables following the generation of suitable rectified prenex normal forms; cf. Example \ref{bsporkpairbsp}, p. \pageref{bsporkpairbsp}.
By virtue of this multiplication of universally quantified expressions, one can avoid the need for
a rule for existential quantifier elimination and the need to decompose the NNFs.
The ability to solve decision problems involving NNFs within the NNF-calculus depends on the extent to which it is possible to define criteria that can be used to decide which universally quantified expressions must be multiplied how often in order
to prove that a formula is a contradiction through the ultimate application of $\forall$E. As we will see, the establishment of such criteria depends on equivalence transformations that include the application of $\wedge$I.

Proving a contradiction by means of unification within the NNF-calculus results in a DNF matrix in which each disjunct contains a literal and its negation.
\begin{defi}
The \emph{DNF matrix} of a formula $\phi$ is the scope that one obtains if $\phi$ is transformed into a prenex normal form and the resulting scope of that prenex normal form is then transformed into a disjunctive normal form (DNF).
\end{defi}

\begin{defi}\label{explizit}
An \emph{explicit} contradiction is a formula without universal quantifiers and with a DNF matrix in which any disjunct contains a literal and its negation.
\end{defi}

In proofs of contradiction in the NNF-calculus, the aim is to derive explicit contradictions by unifying connected literals through the application of $\forall$E subsequent to equivalence transformations that include the application of $\wedge$I.

In addition to $\forall$E and $\wedge$I, the NNF-calculus comprises well-known rules for generating diverse normal forms, such as (i) definitions of sentential connectives and quantifiers as well as De Morgan's laws for generating NNFs, (ii) distributive laws for generating disjunctive and conjunctive normal forms (DNFs and CNFs), and (iii) PN laws (cf. table \ref{pnlaws}) for generating prenex and anti-prenex normal forms. Finally, we presume the laws of substitution for renaming variables (cf. table \ref{subg}), associative and commutative laws for $\wedge$ and $\vee$, and commutative laws for sequences of universal or existential quantifiers. With the exception of $\forall$E, all rules are well-known logical equivalence rules.

\begin{thm}
The NNF-calculus is correct and complete.
\end{thm}

\proof
The correctness of the NNF-calculus follows from the fact that all of its rules are well-known correct derivation rules.
Its completeness can be proven by showing that any proof within the correct and complete tree calculus can be transformed into a proof in the NNF-calculus.
Each quantified expression that is decomposed $n$ times in a proof within the tree calculus is multiplied by applying $\wedge$I $n-1$ times  in the corresponding proof within the NNF-calculus. Subsequently, one generates a prenex normal form in which any quantifier $Y$ that is eliminated later than a quantifier $X$ on the same proof path in the tree calculus is in the scope of $X$. Finally, $\forall$E is applied such that the same literals are unified.
The resulting DNF matrix is contradictory iff all proof paths within the tree calculus are closed.
\qed

In section \ref{undmin}, a further rule, $\exists$M (cf. table \ref{emR}), will be added to the NNF-calculus, but only to avoid applications of $\wedge$I
in proofs of subformulae $\psi$. We will prove that any number of applications of $\exists$M can always be replaced with one application of $\wedge$I when
proving that a subformulae $\psi$ is contradictory; cf. Theorem \ref{undIeinmalth}.

\section{Proofs in the NNF-Calculus}\label{beweisschritte}

This section provides a brief general overview of the essential steps of
proofs of contradiction in the NNF-calculus.  The general strategy for such proofs also underlies the specific decision procedure for subformulae $\psi$ described in the following sections. The specific steps of the proof procedure cannot be explained or justified in detail for the general case of NNFs at this point. In the following sections, the principles of the proof strategy are explained and justified in detail with respect to the special case of subformulae $\psi$.
Terms such as `anti-prenex, optimized FOLDNFs', `$\wedge$I-optimized proof', `$\wedge$I-minimal proof' and `optimal prenex normal forms' are explained in this section only to the extent that is necessary for a rough understanding.

\begin{figure}
	\setlength{\unitlength}{0.8cm}
	\begin{picture}(10,10)\fboxsep2mm
	
	\put(4.5,10){\fbox{NNF $\phi$}}
	
	\put(5.5,9.6){\vector(0,-1){1}}
	
	\put(3.7,7.5){\fbox{\begin{tabular}{l}FOLDNF $\phi^{'}$\\
			$D_{1} \vee \ldots \vee D_{n}$
			\end{tabular}}}
	
	\put(5.5,6.75){\vector(0,-1){1}}
	
	\put(2.25,5){\fbox{anti-prenex, $\wedge$I-optimized $D_{i}^{*}$}}
	
	\put(5.5,4.4){\vector(0,-1){1}}
	
	\put(0,2.6){\fbox{optimal prenex normal form $D_{i}^{**}$ with DNF matrix}}
	
	\put(5.5,2.2){\vector(0,-1){1}}
	
	\put(-0.25,0.5){\fbox{explicit contradiction $D_{i}^{***}$ after the application of $\forall$E}}
	
	\end{picture}
	\caption{Steps of proof in the NNF-calculus \label{chart}}
\end{figure}
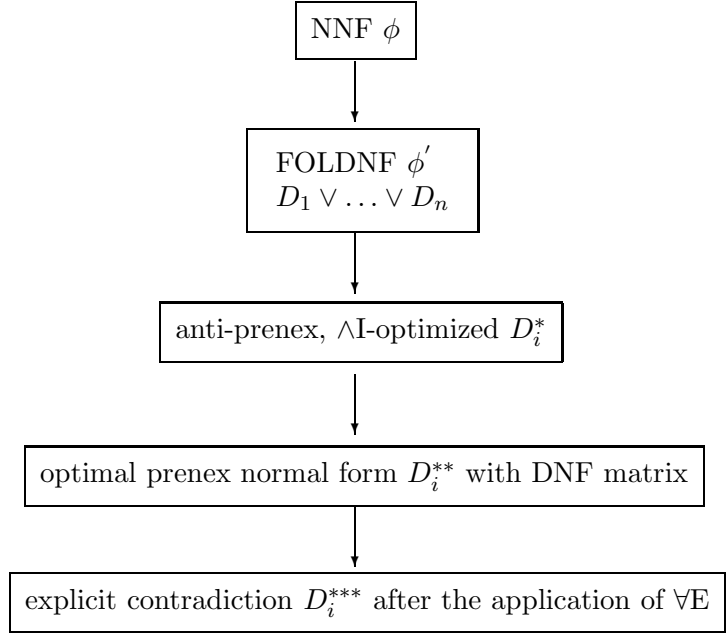

An NNF $\phi$ is first transformed into a sat-equivalent, anti-prenex, optimized FOLDNF $\phi^{'}$ (= $D_{1} \vee \ldots \vee D_{n}$).
An effective procedure for transforming a FOL formula $\phi$ into a logically equivalent FOLDNF $\phi'$ is described in \cite{Lampert}, section 2.
A procedure for optimizing a FOLDNF is introduced %delivered? presented?
in section \ref{optwid} below. It consists of deleting literals from $D_{i}$ that are not part of any unifiable pair of literals in $D_{i}$.
This first step of generating an anti-prenex, optimized FOLDNF drives quantifiers inward to the greatest possible extent by applying PN-laws (= \emph{anti-prenex}, cf. Definition \ref{antiprenex}).
This is essential for the desired proof strategy; cf. section \ref{minscope}. By contrast, $\vee$ is driven outward as far as possible, which makes it possible to reduce the decision problem
for $\phi$ to a decision problem concerning single disjuncts of the resulting FOLDNF.

The second step is to compute, for each single disjunct $D_{i}$, whether a sat-equivalent, anti-prenex, $\wedge$I-optimized formula $D_{i}^{*}$ can be generated.
\emph{$\wedge$I-optimized} means that $\wedge$I is applied no more than is necessary in the case that a %proof by contradiction
proof of contradiction can be identified. I refer to the corresponding proofs as \emph{$\wedge$I-minimal proofs} and to the corresponding proof strategy as the \emph{$\wedge$I-minimal proof strategy}. This step is the most crucial and problematic one. Section \ref{ausblick} briefly summarizes the questions that arise
when computing the applications of $\wedge$I in the general case of FOL formulae. This paper avoids answering these questions by restricting all consideration to $\wedge$-NNFs. It can be shown that in the case of this fragment of FOL, $\wedge$I must be applied no more than once (or even not at all, if one allows for $\exists$M) to prove that a $\wedge$-NNF is contradictory; cf. Theorems \ref{undIeinmalth} and \ref{herbrandth}.

As the third step, each $D_{i}^{*}$ is transformed into an optimal prenex normal form $D_{i}^{**}$ with a DNF matrix. A prenex is \emph{optimal} iff \label{optimalpre1}
the universal quantifier $\forall \nu$ is within the scope of the existential quantifier $\exists \mu$ if $\nu$ is to be replaced with $\mu$ to derive an explicit contradiction through the application of $\forall$E; cf. also Definition \ref{optimal}, p. \pageref{optimal}.

As the last step, $\forall$E is applied to derive an explicit contradiction $D_{i}^{***}$. Each step of the proof is an equivalence transformation with respect to satisfiability.
This is also true in the case of this last, non-essential step because the decision-making is independent of this last step and $\forall$E is applied only if $D_{i}$ is contradictory.

Figure \ref{chart} schematizes proofs in the NNF-calculus in accordance with this proof strategy. Example \ref{bsporkpairbsp} illustrates a $\wedge$I-minimal
proof of contradiction in the NNF-calculus that requires the application of $\wedge$I.

\begin{exa}\label{bsporkpairbsp}
\begin{eqnarray}
& \forall x_{1}(\forall x_{2}\forall  x_{6}Fx_{1}x_{2}x_{6} \vee \forall x_{3}\forall x_{7}\neg Fx_{3}x_{1}x_{7}) \; \wedge & \nonumber \\
& \exists y_{1}\forall x_{4}\exists y_{3}\neg Fy_{1}x_{4}y_{3} \wedge \exists y_{2}\forall x_{5}\exists y_{4}Fx_{5}y_{2}y_{4} & \label{orkpairbsp}
\end{eqnarray}
This is a FOLDNF with only one contradictory disjunct $D_{i}$ that contains $\vee$. The proof depends on the unification of the two pairs of connected literals
$\{Fx_{1}x_{2}x_{6}, \neg Fy_{1}x_{4}y_{3}\}$ and $\{Fx_{5}y_{2}y_{4}, \neg Fx_{3}x_{1}x_{7}\}$. To unify these pairs, $x_{1}$ must be replaced with $y_{1}$ in the first pair and with $y_{2}$ in the second pair. To do so, $\wedge$I must be applied once to duplicate the first conjunct of (\ref{orkpairbsp}).
We prescribe that the variables of quantifiers occurring within the conjuncts resulting from the application of $\wedge$I are to be rectified by increasing the depth of their indices by one. This results in the following formula $D_{i}^{*}$:
\begin{eqnarray}
& \forall x_{1_{1}}(\forall x_{2_{1}}\forall  x_{6_{1}}Fx_{1_{1}}x_{2_{1}}x_{6_{1}} \vee \forall x_{3_{1}}\forall x_{7_{1}}\neg Fx_{3_{1}}x_{1_{1}}x_{7_{1}}) \; \wedge & \nonumber \\
& \forall x_{1_{2}}(\forall x_{2_{2}}\forall  x_{6_{2}}Fx_{1_{2}}x_{2_{2}}x_{6_{2}} \vee \forall x_{3_{2}}\forall x_{7_{2}}\neg Fx_{3_{2}}x_{1_{2}}x_{7_{2}}) \; \wedge &  \label{orkpairbsp2}\\
& \exists y_{1}\forall x_{4}\exists y_{3}\neg Fy_{1}x_{4}y_{3} \wedge \exists y_{2}\forall x_{5}\exists y_{4}Fx_{5}y_{2}y_{4} & \nonumber
\end{eqnarray}

I refer to the number of levels of indices of a variable as its depth. Variables of $D_{i}$ such as $x_{1}$ have a depth of one. Each application of $\wedge$I increases the level of any variables
that are bound by quantifiers occurring in the multiplied conjunct by one. For example, $x_{1_{1}}$ and $x_{1_{2}}$ each have a depth of 2.\label{indexundI}
\begin{defi}\label{derivate}
A \emph{derivate} of a variable $\mu$ of depth 1 is any variable that is identical to $\mu$ if one deletes all indices at levels $>1$.
A \emph{derivate of a pair of literals} $\{L1,L2\}$ with variables of depth 1 is a pair of literals that is identical to $\{L1,L2\}$ if one deletes all indices at levels $>1$.
\end{defi}

In formula (\ref{orkpairbsp2}), $x_{1_{1}}$ and $x_{1_{2}}$ can be replaced with $y_{1}$ and $y_{2}$, respectively, to unify derivates of the original pairs of literals.
However, the application of $\wedge$I and the intended substitutions $x_{1_{1}}/y_{1}$ and $x_{1_{2}}/y_{2}$ cause
the proof to now also depend on the unification of the pair of literals $\{Fx_{1_{2}}x_{2_{2}}x_{6_{2}},$ $\neg Fx_{3_{1}}x_{1_{1}}x_{7_{1}}\}$. This results in the following set of pairs of literals:
\begin{eqnarray}
& \{\{Fx_{1_{1}}x_{2_{1}}x_{6_{1}}, \neg Fy_{1}x_{4}y_{3}\}, & \nonumber\\
& \{Fx_{5}y_{2}y_{4}, \neg Fx_{3_{2}}x_{1_{2}}x_{7_{2}}\}, & \label{orkpairsbsplit}\\
& \{Fx_{1_{2}}x_{2_{2}}x_{6_{2}}, \neg Fx_{3_{1}}x_{1_{1}}x_{7_{1}}\}\} & \nonumber
\end{eqnarray}
The unification of this set of unifiable pairs of literals is sufficient to prove that (\ref{orkpairbsp}) is contradictory.
To unify all pairs of literals in (\ref{orkpairsbsplit}), the $x$ variables must be replaced with $y$ variables in accordance with the following list of substitutions:
\begin{eqnarray}
& \{\{x_{1_{1}},y_{1}\}, \{x_{1_{2}},y_{2}\},\{x_{2_{1}},y_{0}\}, \{x_{2_{2}},y_{1}\}, \{x_{3_{1}},y_{2},\}, \{x_{3_{2}},y_{0}\}, \{x_{4},y_{0}\}, & \nonumber\\
& \{x_{5},y_{0}\}, \{x_{6_{1}},y_{3}\},\{x_{6_{2}},y_{0}\},\{x_{7_{1}},y_{0}\},\{x_{7_{2}},y_{4}\}\} & \label{orkpairsub}
\end{eqnarray}
A prenex normal form $D_{i}^{**}$ with an optimal prenex must be generated from (\ref{orkpairbsp2}). We specify one possible optimal prenex in the form of an ordered list
that contains only the $x$ and $y$ variables:
\begin{eqnarray}
& \{y_{0},y_{1},y_{2},x_{4},y_{3},x_{5},y_{4},x_{1_{1}},x_{1_{2}},x_{2_{1}},x_{2_{2}},x_{3_{1}},x_{3_{2}},x_{6_{1}},x_{6_{2}},x_{7_{1}},x_{7_{2}}\} & \label{orkpairbsppre}
\end{eqnarray}
Formulae (\ref{orkpairbsp}), (\ref{orkpairbsp2}), (\ref{orkpairsbsplit}), (\ref{orkpairsub}), and (\ref{orkpairbsppre}) serve as a recipe that shows how $\wedge$I is to be applied to the $D_{i}$ (\ref{orkpairbsp}) in order to generate the $D_{i}^{*}$ (\ref{orkpairbsp2}) that is to be transformed into a $D_{i}^{**}$ with the optimal prenex (\ref{orkpairbsppre}) in order to unify the literals in (\ref{orkpairsbsplit}) by applying $\forall$E such that an explicit contradiction $D_{i}^{***}$ is obtained.
\end{exa}

The crucial question for the application of the $\wedge$I-minimal proof strategy is in which cases a decision can be made based on computing the minimal necessary applications of $\wedge$I within a finite number of steps.
The proof strategy ultimately serves the purpose of restricting the number of applications of $\wedge$I to the minimum necessary. This makes it possible to terminate a proof path when it is recognized that no more
applications of $\wedge$I can be found on that path that may contribute to a $\wedge$I-minimal proof. Of course, all alternative proof paths that may result in $\wedge$I-minimal proofs must be generated.\footnote{Unlike in the tree calculus, different proof paths search for different $\wedge$I-minimal proofs. Therefore, it is proven that $D_{i}$ is contradictory as soon as a $\wedge$I-minimal proof is found on any proof path. By contrast, in the case that all proof paths terminate without the identification of any $\wedge$I-minimal proof, it is proven that $D_{i}$ is not contradictory.
The question lies in the extent to which such termination can be achieved based on correct criteria for termination; cf. section \ref{ausblick}.}
Decidability depends on whether the search tree can be restricted to a finite number of proof paths of finite depth. According to the $\wedge$I-minimal proof strategy, the computation of the necessary applications of $\wedge$I must be related to purely syntactic criteria. In this paper, we will study this proof strategy only in the rather simple case of subformulae $\psi$ or, consequently, the case of FOLDNFs in which all disjuncts are  $\wedge$-NNFs; cf. p. \pageref{redphitopsi} above.

 \begin{figure}
 	\setlength{\unitlength}{0.8cm}
 	\begin{picture}(10,10)\fboxsep2mm
 	
 	\put(4.5,10){\fbox{NNF $\psi$}}
 	
 	\put(5.5,9.6){\vector(0,-1){1}}
 	
 	\put(3.7,7.5){\fbox{anti-prenex $\psi^{'}$}}
 	
 	\put(5.5,6.75){\vector(0,-1){1}}
 	
 	\put(2.25,5){\fbox{anti-prenex, $\exists$M-optimized $\psi^{''}$}}
 	
 	\put(5.5,4.4){\vector(0,-1){1}}
 	
 	\put(0,2.6){\fbox{optimal prenex normal form $\psi^{*}$ with a DNF matrix}}
 	
 	\put(5.5,2.2){\vector(0,-1){1}}
 	
 	\put(0,0.5){\fbox{explicit contradiction $\psi^{**}$ after the application of $\forall$E}}	
 	\end{picture}
 	\caption{Steps of proof for a subformula $\psi$ \label{chart2}}
 \end{figure}
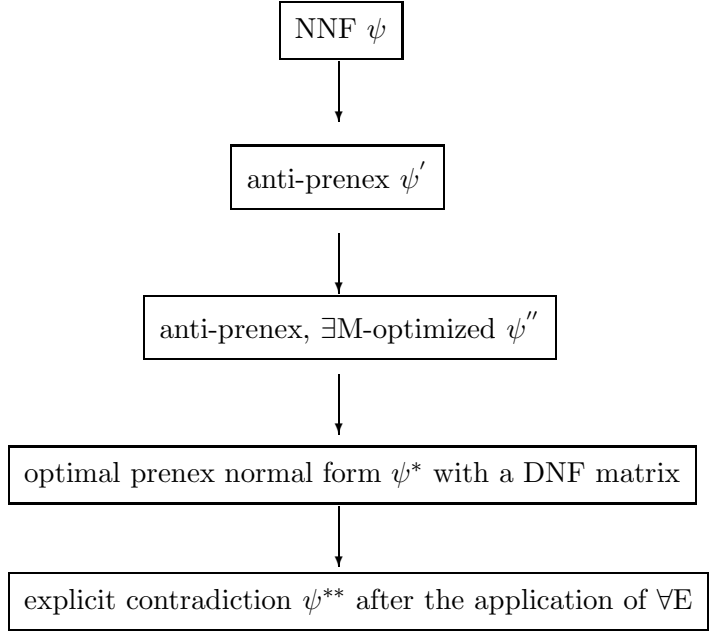

I will abstain from considering FOLDNFs in the following sections \ref{minscope} to \ref{bkrit}; instead, I will consider only subformulae $\psi$. As we will see,
one can decide the refutability of subformulae $\psi$ without any application of $\wedge$I.
I call the resulting sat-equivalent formulae \emph{$\exists$M-optimized} formulae. This terminology and the generation of $\exists$M-optimized formulae are explained in section \ref{eopt}.
Figure \ref{chart2} schematizes proofs for subformulae $\psi$ using $\exists$M in the NNF-calculus.

\section{$\wedge$I-Minimal Proofs}\label{undmin}

We are looking for an intelligent algorithm to decide whether explicit contradictions can be derived in the NNF-calculus and, if so, how.
The proof strategy that will be exemplified in the following for the case of subformulae $\psi$ consists of a systematic search for $\wedge$I-minimal proofs by means of
equivalence transformations.
\begin{defi}
A proof in the NNF-calculus is \emph{$\wedge$I-minimal} iff each application of $\wedge$I in the proof is necessary.
An application of $\wedge$I is \emph{necessary} iff no explicit contradiction could be derived any longer if one of the conjuncts resulting from the application of $\wedge$I were to be eliminated
and all remaining $x$ variables were substituted as before.
\end{defi}
\begin{exa}
Consider the elimination of the second conjunct in formula (\ref{orkpairbsp2}) of Example \ref{bsporkpairbsp}, p.  \pageref{bsporkpairbsp}. Replacing the remaining $x$ variables in accordance with the substitution list given in (\ref{orkpairsub}) does not result in an explicit contradiction. Therefore, the described proof is $\wedge$-minimal.
\end{exa}

Note that $\wedge$I-minimal proofs are defined with respect to given
proofs of contradiction. It is possible for two different $\wedge$I-minimal proofs for the same $D_{i}$ to exist that differ in their numbers of applications of $\wedge$I. A $\wedge$I-minimal proof is therefore not necessarily the proof with the smallest number of applications of $\wedge$I necessary to prove some $D_{i}$ contradictory.

In a $\wedge$I-minimal proof search, $\wedge$I is applied only to unify pairs of literals by means of $\forall$E.
To avoid unnecessary applications of $\wedge$I, one needs to consider only the \emph{minimal} sets of unifiable pairs that are necessary for
proving a contradiction. Likewise, the unification of such a set with substitution list $\sigma1$ must be \emph{minimal}; i.e., no substitution list $\sigma2$ ($\sigma1 \neq \sigma2$) that unifies the same minimal set of unifiable pairs of literals may exist such that any member of $\sigma2$ is a part of a member of $\sigma1$. A $\wedge$I-minimal proof is also incompatible with an application of $\wedge$I that, in turn, \emph{induces} another application of $\wedge$I for the purpose of replacing $x$ variables with $y$ variables in two \emph{similar} conjuncts of $D_{i}^{*}$ by means of \emph{similar} substitutions to unify \emph{similar} pairs of literals; here, conjuncts, substitutions and pairs of literals are considered \emph{similar} iff they are identical after indices at levels $>1$ are deleted (cf. the discussion in section \ref{ausblick} on the `loop criterion').

$\wedge$I-minimal proofs are very elegant; any application of $\wedge$I is the result of a systematic proof search.
%The strategy for searching for $\wedge$I-minimal proofs can similarly be applied to search for the necessary decompositions of quantified expressions in
%proofs of contradiction within the tree calculus.
I will restrict the following discussion to subformulae $\psi$. Deciding upon these formulae allows one to identify unifiable pairs of literals. This capability can be utilized for an optimized search for proofs of contradiction; cf. sections \ref{optwid} and \ref{subkpairs}. Furthermore, subformulae $\psi$ serve as a simple example to illustrate how the $\wedge$I-minimal proof strategy can be used to define a decision procedure. This proof strategy is also interesting because the justification of such a decision procedure does not depend on any model theory or semantics; it is sufficient to refer to syntactic considerations concerning the unification of pairs of literals within the NNF-calculus.

Finally, a $\wedge$I-minimal proof strategy is also interesting because it is efficient, in a certain sense.
Proofs of contradiction within the tree calculus become inefficient when quantified expressions are decomposed several times without each decomposition being necessary for the proof.
In the NNF-calculus, the same applies to applications of $\wedge$I for the replacement of $x$ variables with $y$ variables that do not contribute to a
proof of contradiction.
$\wedge$I-minimal proofs restrict the elimination of universal quantifiers to the minimum extent necessary for unifying the members of a minimal set of pairs of literals.
I will not discuss questions of efficiency any further in this paper because from here on, I will consider only simple subformulae $\psi$ to show that a decision procedure for these formulae is \emph{optimally $\wedge$I-minimal}.

\begin{defi}
A decision procedure for formulae $\phi$ is \emph{optimally $\wedge$I-minimal} if it is possible to decide whether the formulae $\phi$ are contradictory without applying $\wedge$I.
\end{defi}

In a certain respect, optimally $\wedge$I-minimal decision procedures correspond to decision procedures with blocking in the tree calculus.\footnote{Blocking restricts the introduction of new variables
due to existential quantifier elimination in the tree calculus.} In general, $\wedge$I is necessary to prove that formulae are contradictory, as illustrated by Example \ref{bsporkpairbsp}.
In the case of subformulae $\psi$, however, one can abstain from the application of $\wedge$I altogether. If $\psi$ is contradictory, then an optimally $\wedge$I-minimal decision procedure for $\psi$ will find a sat-equivalent formula $\psi^{*}$ that allows one to unify the pair of literals in $\psi^{*}$ by applying $\forall$E without applying $\wedge$I at all. If $\psi$ is not contradictory, then computation of the minimal applications of $\wedge$I prior to any actual application of either $\wedge$I or $\forall$E will reveal that no such sat-equivalent formula $\psi^{*}$ can be found.
In the following, I will define such an equivalence transformation that will result in a formula $\psi^{*}$ without the need to apply $\wedge$I in the case of a contradictory formula $\psi$.
The ability to avoid the application of $\wedge$I is a significant feature of subformulae $\psi$. The ideal of an optimally $\wedge$I-minimal decision procedure can be realized in the case of such formulae.

To make this possible, however, one must use the following rule under certain conditions:
\begin{longtable}{rcll}
$\exists \mu (\varphi(\mu) \wedge \psi(\mu))$ & $\vdash$ & $\exists \mu_{1}\varphi(\mu/\mu_{1}) \wedge \exists \mu_{2}\psi(\mu/\mu_{2})$ & $\exists \mbox{M}$ \label{emR}\\
\caption{Existential Quantifier Multiplication}
\end{longtable}
We allow for indices of $y$ variables at levels $>1$ subsequent to the application of $\exists$M. %By renaming, however, one can generate rectified formula with depth 1 anytime.

When an existential quantifier is multiplied by $\exists$M, the variable bound by the multiplied quantifier is `separated' within the scope of the quantifier.
\begin{defi}
A variable is \emph{separated} if it is subscripted with different indices at different positions subsequent to the application of $\exists$M or other logical rules.
\end{defi}
Hence, the application of $\exists$M separates $\mu$ in the scope of $\exists \mu (\varphi(\mu) \wedge \psi(\mu))$ such that the occurrences of $\mu$ are subscripted with an index of 1 in the first conjunct and with an index of 2 in the second conjunct.

$\exists$M can be applied at any level of the logical hierarchy  of NNFs, like all other rules of the NNF-calculus.
This rule must be applied in an optimally $\wedge$I-minimal decision procedure for subformulae $\psi$.
Such applications, however, must be restricted to those that are sat-equivalent. Section \ref{eopt} specifies under what conditions this is the case. Section \ref{em1} proves that the refutability of subformulae $\psi$ can be decided within a $\wedge$I-minimal proof strategy without $\exists$M by applying $\wedge$I no more than once. First, however, the following section \ref{minscope} explains why and how the scopes in $\psi$ must be minimized in the $\wedge$I-minimal proof strategy.

\section{Minimizing Scopes}\label{minscope}

Given a subformula $\psi$ in prenex normal form, the following conditions make it impossible to unify a pair of connected literals $\{L1,L2\}$ in $\psi$ through the direct application of $\forall$E without first applying equivalence transformations (that include $\wedge$I or $\exists$M):
\begin{description}
\item[suboptimal prenex] An $x$ variable $\nu$ must be replaced with a $y$ variable $\mu$ to unify $\{L1,L2\}$, but $\exists \mu$ is in the scope of $\forall \nu$ in $\psi$.\label{suboptimalprae}
\item[$x$ ambiguity] An $x$ variable $\nu$ must be replaced with (at least) two existential $y$ variables to unify $\{L1,L2\}$.
\label{xueindeutigkeit}
\end{description}
One cannot straightforwardly infer whether $\psi$ is contradictory under these conditions. However, this situation changes if one considers equivalence transformations, in which minimizing scopes as much as possible is crucial.

\begin{longtable}{rcll}
$\forall \nu (A \, \wedge \, B(\nu))$ &  $\dashv\vdash$ &   $A \, \wedge \, \forall \nu B(\nu)$& PN1\\
$\forall \nu (B(\nu) \, \wedge \, A)$ & $\dashv\vdash$ &  $\forall \nu B(\nu) \, \wedge \, A$& PN2\\
$\forall \nu (A \, \vee \, B(\nu))$ &  $\dashv\vdash$ &   $A \, \vee \, \forall \nu B(\nu)$& PN3\\
$\forall \nu (B(\nu) \, \vee \, A)$  & $\dashv\vdash$ & $\forall \nu B(\nu) \, \vee \, A$& PN4\\
$\exists \nu (A \, \wedge \, B(\nu))$ &  $\dashv\vdash$ &  $A \, \wedge \, \exists \nu B(\nu)$ & PN5\\
$\exists \nu (B(\nu) \, \wedge \, A)$ & $\dashv\vdash$ &  $\exists \nu B(\nu) \, \wedge \, A$& PN6\\
$\exists \nu (A \, \vee \, B(\nu))$ &  $\dashv\vdash$ &   $A \, \vee \, \exists \nu B(\nu)$& PN7\\
$\exists \nu (B(\nu) \, \vee \, A)$ & $\dashv\vdash$ &  $\exists \nu B(\nu) \, \vee \, A$& PN8\\
$\forall \nu (A(\nu) \, \wedge \, B(\nu))$ & $\dashv\vdash$ &  $\forall \nu A(\nu) \, \wedge \, \forall \nu B(\nu)$& PN9\\
$\exists \nu (A(\nu) \vee B(\nu))$  &  $\dashv \vdash$ &   $\exists \nu A(\nu) \vee \exists \nu B(\nu)$& PN10\\
\caption{PN Laws}\label{pnlaws}
\end{longtable}

\begin{defi}\label{antiprenex}
\emph{Anti-prenex normal forms} are NNFs with quantifiers that have been driven inward as far as possible through the application of PN laws; cf. table \ref{pnlaws}.\footnote{
Cf. \cite{Lampert}, section 2, for the history of anti-prenex normal forms and various algorithms for minimizing the scopes of quantifiers (such as described in \cite{Nonnengart}) or \cite{Quine}, p.126-129).}
\end{defi}
Anti-prenex normal forms are a crucial part of a $\wedge$I-minimal proof strategy to enable the computation of all possible \emph{optimized prenex normal forms}; cf. section \ref{wpopt}.

\begin{defi}\label{optimiert}
An \emph{optimized prenex normal form} of NNF $\phi$ is a prenex normal form of $\phi$ in which the universal quantifiers have been brought within the scope of the existential quantifiers to the greatest possible extent by applying PN laws to generate prenex normal forms from an anti-prenex normal form of $\phi$.
\end{defi}
Optimized prenex normal forms are important for deciding whether there exists some \emph{optimal} prenex, in which a universal quantifier $\forall \nu$ is within the scope of an existential quantifier $\exists \mu$ if $\nu$ is to be replaced with $\mu$ for unification; cf. p. \pageref{optimalpre1} above and Definition \ref{optimal} below.

Furthermore, within the $\wedge$I-minimal proof strategy, generating anti-prenex normal forms is important for multiplying universal quantifiers by applying PN9 to the maximal extent.\footnote{In the case of anti-prenex forms that contain $\vee$, the scope of universal quantifiers must be transformed into a CNF, and the scope of existential quantifiers must be transformed into a DNF under certain conditions. This results in further multiplications of quantifiers; cf. \cite{Lampert}, section 2. In the case of a decision procedure for $\wedge$-NNFs, however, we can avoid such transformations.} This avoids the need to apply $\wedge$I in the case that $x$ variables must be replaced with different $y$ variables in different literals for unification.
Minimizing scopes is a strategy for computing how many applications of $\wedge$I, if any, are necessary
to generate optimal prenexes and to replace $x$ variables in different literals with different $y$ variables.
Applications of $\wedge$I are at most necessary if, after transformation into anti-prenex normal forms, no optimal prenex can be generated or an $x$ variable must be replaced with more than one $y$ variable for unification.

Because $\psi$ does not contain $\vee$, it is sufficient to apply PN1, PN2, PN5, PN6 and PN9 to generate the anti-prenex normal form $\psi'$ from $\psi$ by means of the following algorithm:

\begin{algo} \label{malg} \hfill
\begin{enumerate}
\item $\psi = \psi'$.
\item If $\psi'$ contains a subformula $\rho$ in the form of $\forall \nu (A \, \wedge \, B(\nu))$, apply PN1 and replace $\rho$ in $\psi'$ with $A \, \wedge \, \forall \nu B(\nu)$. The new result becomes $\psi'$. Go to (2) if the new result $\psi'$ is different from the initial formula $\psi'$; otherwise, go to (3).
\item If $\psi'$ contains a subformula $\rho$ in the form of $\forall \nu (B(\nu) \, \wedge \, A)$, apply PN2 and replace $\rho$ in $\psi'$ with $\forall \nu (B(\nu) \, \wedge \, A)$.
The new result becomes $\psi'$. Go to (2) if the new result $\psi'$ is different from the initial formula $\psi'$; otherwise, go to (4).
\item If $\psi'$ contains a subformula $\rho$ in the form of $\exists \nu (A \, \wedge \, B(\nu))$, apply PN5 and replace $\rho$ in $\psi'$ with $A \, \wedge \, \exists \nu B(\nu)$.
The new result becomes $\psi'$. Go to (2) if the new result $\psi'$ is different from the initial formula $\psi'$; otherwise, go to (5).
\item If $\psi'$ contains a subformula $\rho$ in the form of $\exists \nu (B(\nu) \, \wedge \, A)$, apply PN6 and replace $\rho$ in $\psi'$ with $\exists \nu B(\nu) \, \wedge \, A$.
The new result becomes $\psi'$. Go to (2) if the new result $\psi'$ is different from the initial formula $\psi'$; otherwise, go to (6).
\item If $\psi'$ contains a subformula $\rho$ in the form of $\forall \nu (A(\nu) \, \wedge \, B(\nu))$, apply PN9 and replace $\rho$ in $\psi'$ with $\forall \nu A(\nu) \, \wedge \, \forall \nu B(\nu)$. The new result becomes $\psi'$. Go to (2) if the new result $\psi'$ is different from the initial formula $\psi'$; otherwise, go to (7).
\item Maximally subscript $\psi'$, return the rectified result $\psi'$, and terminate.
\end{enumerate}
\end{algo}
When Algorithm \ref{malg} reaches (7), either no quantifier is present above $\wedge$ in $\psi'$ or $\psi'$ contains a subformula $\rho$ in the form of  $\exists \nu (B(\nu) \, \wedge \, A(\nu))$.

%\newpage

\begin{thm}\label{mlog}
Algorithm \ref{malg} is a logical equivalence transformation.
\end{thm}
\proof
Algorithm \ref{malg} involves nothing but the application of PN laws (cf. table \ref{pnlaws}) and laws of substitution (cf. table \ref{subg}). These laws are all logical equivalence rules.\qed
Since Algorithm \ref{malg} is a logical equivalence transformation, $\psi'$ is a forteriori sat-equivalent to $\psi$.

\begin{thm}\label{mterm}
Algorithm \ref{malg} terminates.
\end{thm}
\proof
This theorem follows from the fact that each rule reduces the depth of the formula, starting with a quantifier.
If $\psi$ contains $n$ quantifiers, the algorithm will terminate after no more than $n$ applications of one of the PN laws and $2\cdot n$ subsequent applications of SUB1 or SUB2.\qed

Let $\psi'$ contain a sequence of existential quantifiers of length $>1$ above $\wedge$, let the innermost existential quantifier of that sequence bind a variable that occurs in both conjuncts, and let the same not hold for all existential quantifiers of that sequence. In this case, sorting the quantifiers of the sequence such that all existential quantifiers that bind a variable occurring in only one conjunct are to the right of all other existential quantifiers in the sequence makes it possible to drive those quantifiers farther inward by applying PN5 or PN6.
However, there is no need to consider such quantifier sorting here because it does not allow for any further application of PN1, PN2 or PN9, and it is only the application of these rules that leads to the intended effect of universal quantifiers either no longer being in the scope of certain existential quantifiers or being multiplied.

\begin{exa}\hfill
\renewcommand{\arraystretch}{1.3}
\begin{longtable}{|l|c|l|}\hline
\bf{No.} & \bf{Formula} & \bf{Rule}\\\hline
(1) & $\exists y_{1}\forall x_{1}\exists y_{2}\forall x_{2}\exists y_{3}(Fx_{1}x_{2}y_{1}y_{1} \wedge \neg Fy_{2}y_{3}x_{1}y_{1})$ & $\psi$\\\hline
(2) &  $\exists y_{1}\forall x_{1}\exists y_{2}\forall x_{2}(Fx_{1}x_{2}y_{1}y_{1} \wedge \exists y_{3}\neg Fy_{2}y_{3}x_{1}y_{1})$ & PN5\\\hline
(3) &  $\exists y_{1}\forall x_{1}\exists y_{2}(\forall x_{2}Fx_{1}x_{2}y_{1}y_{1} \wedge \exists y_{3}\neg Fy_{2}y_{3}x_{1}y_{1})$ & PN2\\\hline
(4) &  $\exists y_{1}\forall x_{1}(\forall x_{2}Fx_{1}x_{2}y_{1}y_{1} \wedge \exists y_{2}\exists y_{3}\neg Fy_{2}y_{3}x_{1}y_{1})$ & PN5\\\hline
(5) &  $\exists y_{1}(\forall x_{1}\forall x_{2}Fx_{1}x_{2}y_{1}y_{1} \wedge \forall x_{1}\exists y_{2}\exists y_{3}\neg Fy_{2}y_{3}x_{1_{2}}y_{1})$ & PN9\\\hline
(6) &  $\exists y_{1}(\forall x_{1}\forall x_{2}Fx_{1}x_{2}y_{1}y_{1} \wedge \forall x_{3}\exists y_{2}\exists y_{3}\neg Fy_{2}y_{3}x_{3}y_{1})$ & SUB2\\\hline
\caption{Transformation in Accordance with Algorithm \ref{malg}}
\end{longtable}
\end{exa}

In step 1 of the sat-equivalence transformation that makes it possible to decide the refutability of $\psi$, the subformula $\psi$ is transformed into an anti-prenex formula $\psi'$.
However, this is not sufficient to define a sat-equivalence transformation that drives quantifiers inward to the maximal extent. For this purpose, $\exists$M must be applied to drive quantifiers above $\wedge$ inward if they bind variables in both conjuncts and if such an application is sat-equivalent. The following section specifies the conditions for a sat-equivalent application of $\exists$M.

\section{$\exists$M-Optimization}\label{eopt}

Applying $\exists \mbox{M}$ enables one to apply PN laws to further minimize the scopes of quantifiers.
\begin{defi}
\emph{$\exists$M-optimization} is a sat-equivalent application of $\exists$M to an anti-prenex, rectified formula $\psi'$ followed by the application of Algorithm \ref{malg}.
\end{defi}
The initial formula $\psi$ must be transformed into a sat-equivalent, anti-prenex and $\exists$M-optimized formula $\psi''$ to
achieve proofs of contradiction without the need to apply $\wedge$I.

In the following, I will first identify two cases in which the application of $\exists$M is not sat-equivalent. Then, I will prove that these are the only two cases in which the application of $\exists$M is not sat-equivalent. In doing so, I will refer to nothing but the question of the unifiability of $\{L1,L2\}$ in $\psi'$. Thus, the sat-equivalent applications of $\exists$M will be specified without referring to model theory or semantics.

What I call the `direct case' of the violation of the requirements for a sat-equivalent application of $\exists$M is defined below.
\begin{defi}\label{dFall}
\emph{Direct Case}: The $y$ variable $\mu$ that is to be multiplied and, thus, separated by applying $\exists$M occurs in identical positions in L1 and L2 before $\exists$M is applied.
\end{defi}
In this case, separating a $y$ variable that occurs in identical positions before the application of $\exists$M transforms a pair of connected literals into a pair of unconnected literals, which is no longer unifiable.  This is already true in the simplest case of $\exists y_{1}(Fy_{1} \wedge \neg Fy_{1})$. This formula is contradictory.
Applying $\exists$M leads to the satisfiable (non-contradictory) formula $\exists y_{1_{1}} Fy_{1_{1}} \wedge \exists y_{1_{2}} \neg Fy_{1_{2}}$.
Therefore, the application of $\exists$M is manifestly not sat-equivalent in the direct case.
In this case, `$y$ unambiguity' is lost:
the same $y$ variable occurs in identical positions before the application of $\exists$M,
whereas different $y$ variables occupy these positions after the application of $\exists$M. Hence, we exclude such cases from among the valid applications of $\exists$M.

The second case in which the requirements for a sat-equivalent application of $\exists$M are violated is what I call the `indirect case'. To define it, let us introduce the concept of an `$xx$ list', which we define recursively:
\begin{defi} The \emph{$xx$ lists} of a pair of connected literals $\{L1,L2\}$ are generated as follows:
\begin{enumerate}\label{defxxlist}
\item If an $x$ variable $\nu_{1}$ and an $x$ variable $\nu_{2}$ occur in identical positions in L1 and L2, then $\nu_{1}$ and $\nu_{2}$ are members of the same $xx$ list.
\item If the $x$ variables $\nu_{1}$ and $\nu_{2}$ are members of the same $xx$ list and if $\nu_{2}$ and the $x$ variable $\nu_{3}$ are also members of the same $xx$ list, then $\nu_{1}$, $\nu_{2}$ and $\nu_{3}$ are all members of the same $xx$ list.
\item (1) and (2) are the only conditions that determine members of the same $xx$ list.
\end{enumerate}
\end{defi}
All $x$ variables of an $xx$ list must be replaced with the same $y$ variable to unify $\{L1,L2\}$.

\begin{exa}
Let $\psi'$ be the following formula:
\begin{eqnarray}
& \exists y_{1}(\forall x_{1}\forall x_{2}Fy_{1}x_{1}x_{1}x_{2}x_{2} \wedge \forall x_{3}\forall x_{4}\neg Fx_{3}x_{3}x_{4}x_{4}y_{1}) & \label{indirekt2}
\end{eqnarray}
The following pair of connected literals $\{L1,L2\}$ must be unified:
\begin{eqnarray}
& \{Fy_{1}x_{1}x_{1}x_{2}x_{2}, \neg Fx_{3}x_{3}x_{4}x_{4}y_{1}\}. & \label{indirekt2lit}
\end{eqnarray}
$x_{1}$ occurs in the same positions as $x_{3}$ and $x_{4}$. Therefore, $x_{1}$ and $x_{3}$ as well as $x_{1}$ and $x_{4}$ are members of the same $xx$ list, according to
Definition \ref{defxxlist}(1). Hence, $x_{1}$, $x_{3}$ and $x_{4}$ are also members of the same $xx$ list, according to Definition \ref{defxxlist}(2).
$x_{2}$ is in the same position as $x_{4}$. Therefore, $x_{2}$ and $x_{4}$ are members of the same $xx$ list, according to Definition \ref{defxxlist}(1). Since $x_{4}$
is also a member of the $xx$ list $\{x_{1},x_{3},x_{4}\}$, $x_{2}$ is a member of this list as well, according to Definition \ref{defxxlist}(2).
This results in the following $xx$ list, which is the only $xx$ list of (\ref{indirekt2lit}):
\begin{eqnarray}
& \{x_{1},x_{2},x_{3},x_{4}\} & \label{indirekt2xx}
\end{eqnarray}
\end{exa}
%\begin{exa}
%The $xx$-lists of the following pair of connected literals
%\begin{eqnarray}
%& \{Fx_{1}x_{2}x_{3}y_{2}x_{5}x_{5}x_{7},\neg Fx_{3}x_{1}x_{4}x_{4}x_{6}y_{1}y_{2}\} & \label{xxlbeisp}
%\end{eqnarray}
%are the following:
%\begin{enumerate}
%\item $\{x_{1},x_{2},x_{3},x_{4}\}$,
%\item $\{x_{5},x_{6}\}$.
%\end{enumerate}
%\end{exa}
%

Furthermore, we define $xy$ and $yx$ pairs as follows:
\begin{defi}
An \emph{$xy$ pair} is an ordered list $\{xvar,yvar\}$ consisting of an $x$ variable $xvar$ and a $y$ variable $yvar$, where $xvar$ occurs in L1 in a position $n$ and $yvar$ occurs in L2 in the same position $n$.
A \emph{$yx$ pair} is an ordered list $\{yvar,xvar\}$ consisting of a $y$ variable $yvar$ and an $x$ variable $xvar$, where $yvar$ occurs in L1 in a position $n$ and $xvar$ occurs in L2 in the same position $n$.
\end{defi}
An $x$ variable $xvar$ must be replaced with the $y$ variable $yvar$ if $xvar$ and $yvar$ occur in the same $xy$ or $yx$ pair.

%\begin{exa}
%$\{x_{5},y_{1}\}$ is a $xy$-pair and $\{y_{2},x_{4}\}$ a $yx$-pair of (\ref{xxlbeisp}).
%\end{exa}
\begin{exa}
(\ref{xyindirekt2}) is the only $xy$ pair of (\ref{indirekt2lit}), and (\ref{yxindirekt2}) is the only $yx$-pair of (\ref{indirekt2lit}).
\begin{eqnarray}
& \{x_{2},y_{1}\} & \label{xyindirekt2}\\
& \{y_{1},x_{3}\} & \label{yxindirekt2}
\end{eqnarray}
\end{exa}
(\ref{indirekt2}) is contradictory: (\ref{indirekt2lit}) can be unified by replacing all $x$ variables with $y_{1}$, which is possible because $\{y_{1},x_{1},x_{2},x_{3},x_{4}\}$ represents an optimal prenex that can be generated by applying PN laws to (\ref{indirekt2}).
The application of $\exists$M to (\ref{indirekt2}), however, is not sat-equivalent because this would separate $y_{1}$ into $y_{1_{1}}$ in the first conjunct and $y_{1_{2}}$ in the second conjunct.
Consequently, $x_{3}$ must be replaced with $y_{1_{1}}$ because of (\ref{yxindirekt2}), and $x_{2}$ must be replaced with $y_{1_{2}}$ because of (\ref{xyindirekt2}).
However, according to (\ref{indirekt2xx}), $x_{1}$, $x_{2}$, $x_{3}$ and $x_{4}$ all must be replaced with the same $y$ variable to unify (\ref{indirekt2lit}).

Based on the above discussion, the indirect case of the violation of the requirements for a sat-equivalent application of $\exists$M can be defined as follows:
\begin{defi}
\emph{Indirect case}: $\{L1,L2\}$ has an $xx$ list $A$ such that at least one $x$ variable $xvar1$ from $A$ is a member of an $xy$ list $\{xvar1,yvar\}$,
at least one $x$ variable $xvar2$ from $A$ ($xvar1 \neq xvar2$) is a member of a $yx$ list $\{yvar,xvar2\}$, and all universal quantifiers binding the $x$ variables in $A$
are within the scope of $\exists yvar$ in $\psi$, where $yvar$ is the $y$ variable multiplied by $\exists$M.
\end{defi}
On p. \pageref{case3beta} below, I will justify why this case is restricted to $x$ variables that are bound by quantifiers in the scope of the existential quantifier that is multiplied by $\exists$M (Case 3($\beta$) below).
In contrast to the direct case, one might say that `$x$ unambiguity' is lost in the indirect case: whereas all $x$ variables of an $xx$ list must be replaced with one and the same $y$ variable before
the application of $\exists$M, they must be replaced with different $y$ variables after the application of $\exists$M. Hence, we exclude not only the direct case but also the indirect case from among the valid applications of $\exists$M.

In the following, I justify why the direct and indirect cases are the only two cases that must be excluded to restrict $\exists$M to sat-equivalent applications.
In doing so, we consider nothing but logically valid equivalence transformations to unify positions of a pair of connected literals within a subformula $\psi'$.

The syntactic properties that determine the unifiability of a pair of connected literals are (i) the order of the quantifiers and (ii) the identity or difference of the variables at certain positions of the literals. The order of the quantifiers, however, is not adversely affected by the application of $\exists$M because any $\exists$M-optimization results in a minimized scope.
This can only increase the possibility of generating optimal prenexes. In addition to the changes to quantifier order, $\exists$M separates the multiplied $y$ variable as well as those $x$ variables that (i) are bound by a universal quantifier that is to the left of the multiplied existential quantifier before $\exists$M is applied and (ii) occur in both literals, L1 and L2.
The separation of $x$ variables subsequent to the separation of the multiplied $y$ variable is due to the application of PN9 in $\exists$M-optimization. This separation
cannot have a negative effect on unifiability because it enables the coverage of positions in different literals with different $y$ variables.
Therefore, the unifiability of a pair of literals can only be adversely affected by the separation of the multiplied $y$ variable, which is subscripted
differently in the two conjuncts subsequent to $\exists$M-optimization.

To identify the cases in which the separation of the multiplied $y$ variable $\mu$
transforms a unifiable pair of connected literals into a pair of literals that is no longer unifiable, we distinguish three cases:
\begin{description}
\item[Case 1] $\mu$ occurs in identical positions in L1 and L2.
\item[Case 2] $\mu$ occurs in identical positions to those of $x$ variables that are not members of an $xx$ list.
\item[Case 3] $\mu$ occurs in identical positions to those of $x$ variables that are members of an $xx$ list.
\end{description}
Note that there is no need to distinguish a case in which $\mu$ occurs in the same position as some $y$ variable that is not identical to $\mu$ because this violates condition (4) of the definition of a pair of connected literals; cf. Definition \ref{konnektiert}, p. \pageref{konnektiert}. Therefore, these three cases are exhaustive. It may happen that the conditions for more than one of the three cases are satisfied for one and the same pair of connected literals. However, the three cases can be considered separately because  the application of $\exists$M is not sat-equivalent iff the multiplication of $\mu$ does not preserve unifiability in regard to at least one case.

Case 1 is the direct case. Case 2 cannot induce applications of $\exists$M that are not sat-equivalent, as illustrated by the following reasoning: The $x$ variables need to be replaced with $\mu$ either ($\alpha$) in only one of
the two literals or ($\beta$) in both. In the first case ($\alpha$), separation of the $y$ variable cannot adversely affect the possibility of unifying the positions in which the $x$ variables occur because the number of $y$ variables with which those $x$ variables need to be replaced for unification does not change. The second case ($\beta$) requires that the universal quantifiers that bind $x$ variables that occur in identical positions to those of $\mu$ must be to the left of $\exists \mu$ before the application of $\exists$M because otherwise, the $x$ variables will not occur in both L1 and L2 due to the presumed maximal subscription. However, given that (a) $\exists \mu$ is in the scope of a universal quantifier $\forall \nu$ in the rectified formula $\psi'$ and (b) $\nu$ and $\mu$ occur in both L1 and L2 in identical positions (and $\nu$, therefore, must be replaced with $\mu$ in both literals), L1 and L2 are not unifiable anyway, and consequently, $\psi'$ is not contradictory. $\psi'$ is not contradictory because it is impossible in case ($\beta$) to generate an optimal prenex that would permit the necessary substitutions, even after the minimization of scopes through $\exists$M-optimization or the application of $\wedge$I (which of the two is done makes no difference; cf. section \ref{em1}). In any case, at least one universal quantifier %$\forall \nu$ or
$\forall \nu_{i}$ ($\nu_{i}$ = $\nu$ with an additional subscript of 1 or 2) remains that is necessarily not in the scope of %$\exists \mu$ or 
$\exists \mu_{j}$ ($\mu_{i}$ = $\mu$ with an additional subscript of 1 or 2) in any optimized prenex.

The indirect case is a special case of case 3. Again, one can ignore a case ($\alpha$) in which there is either no $xy$ pair that contains both an $x$ variable from an $xx$ list and $\mu$ as members or no $yx$ pair that contains both an $x$ variable from an $xx$ list and $\mu$ as members. In such a case, the number of necessary substitutions in regard to the $x$ variables of the $xx$ list would again not be increased by applying $\exists$M. However, in a case ($\beta$) in which some $x$ variable $xvar1$ in an $xx$ list occurs together with $\mu$ in an $xy$ pair and some $x$ variable $xvar2$ in the same $xx$ list occurs together with $\mu$ in a $yx$ pair, the situation is different. Nevertheless, we can identify several conditions under which the application of $\exists$M is still sat-equivalent. First of all, we can assert that $xvar1 \neq xvar2$ because otherwise the case reduces to case 2. 
\label{case3beta}Furthermore, if one universal quantifier $\forall \nu$ that binds an $x$ variable $\nu$ from the $xx$ list is to the left of $\exists \mu$ in $\psi'$, then $\{L1,L2\}$ is not unifiable, and $\psi'$ is not contradictory. This is so because %given that (a) $\exists \mu$ is in the scope of $\forall \nu$ in the rectified formula $\psi'$ and (b) $\nu$ occurs in an $xx$ list and the conditions of case 3($\beta$) are satisfied, 
in this case, minimizing scopes through $\exists$M-optimization or the application of $\wedge$I is necessary. However, minimizing scopes or applying $\wedge$I, where both include maximal subscription, makes it necessary for at least one $x$ variable from the $xx$ list to be replaced with both $\mu_{1}$ and $\mu_{2}$ for unification in one and the same literal, which is logically impossible. Therefore, the only subcase of case 3 is the indirect case, in which $xvar1 \neq xvar2$ and all universal quantifiers binding $x$ variables from an $xx$ list are in the scope of $\exists \mu$ in $\psi'$.
%\label{case3beta}Furthermore, if one universal quantifier $\forall \nu$ that binds an $x$ variable $\nu$ from the $xx$ list is to the left of $\exists \mu$ in $\psi'$, then $\{L1,L2\}$ is not unifiable, and $\psi'$ is not contradictory. Again, this is so because it is impossible in case ($\beta$) to generate an optimal prenex that would permit the necessary substitutions, even after the minimization of scopes through $\exists$M-optimization or the application of $\wedge$I (which of the two is done makes no difference; cf. section \ref{em1}). In any case, either $\forall \nu_{1}$ had to be replaced with $\exists \mu_{1}$ or $\forall \nu_{2}$ with $\exists \mu_{2}$. As $\forall \nu_{1} \ldots \exists \mu_{1}$ or $\forall \nu_{2} \ldots \exists \mu_{2}$ occur in the same conjunct, the existential quantifier is necessarily in the scope of the universal quantifier in any optimized prenex.
%Therefore, the only subcase of case 3 is the indirect case, in which $xvar1 \neq xvar2$ and all universal quantifiers binding $x$ variables from an $xx$ list are in the scope of $\exists \mu$ in $\psi'$.

Thus, the direct and indirect cases are the only cases that must be excluded to restrict the application of $\exists$M to sat-equivalent transformations.
\begin{defi}
$\exists$M$_{B}$ is the rule stating that $\exists$M is applied unless the conditions for either the direct or the indirect case are satisfied.
\end{defi}

\begin{thm}\label{ebt}
The application of $\exists$M$_{B}$ is sat-equivalent.
\end{thm}

\proof
This theorem follows from the above distinction of cases, which shows that the direct and indirect cases are the only cases that imply an application of $\exists$M that is not sat-equivalent.
The proof is based on nothing but proof-theoretic considerations concerning the unification of the pair of connected literals in $\psi'$ in the NNF-calculus.
\qed

The following algorithm transforms the initial subformula $\psi$ into a sat-equivalent, anti-prenex and $\exists$M-optimized subformula $\psi''$:

\begin{algo}\hfill \label{mem}
\begin{enumerate}
\item Apply Algorithm \ref{malg} to $\psi$. The result is $\psi'$.
\item If an existential quantifier exists above $\wedge$, apply $\exists$M$_{B}$ to multiply this quantifier if possible.
The result is $\psi''$.
\item If $\psi' = \psi''$, return $\psi''$ and terminate; otherwise, go back to (1).
\end{enumerate}
\end{algo}

\begin{thm}\label{memsat}
Algorithm \ref{mem} is a sat-equivalence transformation.
\end{thm}

\proof
This theorem follows from Theorems \ref{mlog} and \ref{ebt}.
\qed

\begin{thm}\label{memterm}
Algorithm \ref{mem} terminates.
\end{thm}

\proof
This theorem follows from Theorem \ref{mterm} and the fact that there can be only a finite number of existential quantifiers in the logical hierarchy above $\wedge$ in $\psi$.
\qed

\begin{thm}\label{minmax}
Algorithm \ref{mem} minimizes the scope of universal quantifiers in $\psi$ to a maximal extent under equivalence transformations and thus maximizes the number of universal quantifiers binding the variables occurring in a pair of connected literals to a maximal extent under equivalence transformations.
\end{thm}

\proof
The scope of universal quantifiers is not minimized any further by Algorithm \ref{mem} only if an existential quantifier exists that is in their scope and above $\wedge$ and if the multiplication of this quantifier is not sat-equivalent. Minimizing scopes maximizes the number of applications of PN9 and, therefore, the number of universal quantifiers.
\qed

Either the resulting formula $\psi''$ is a conjunction of two quantified expressions, each one containing exactly one literal,
or $\psi''$ is a quantified expression with an outermost sequence of quantifiers ending with an existential quantifier $\exists \mu$ above $\wedge$, where each conjunct contains exactly one literal and the conditions for either the direct or the indirect case are satisfied.

\section{$\exists$M-Optimization and the Application of $\wedge$I}\label{em1}

$\exists$M-optimization minimizes scopes and separates variables. This makes applications of $\wedge$I superfluous in the case of subformulae $\psi'$.
This section demonstrates this claim, first by example and then in general.

\begin{exa}
Let $\psi'$ be the following formula:

\begin{eqnarray}
& \forall x_{1} \exists y_{1}(\forall x_{3} Fy_{1}x_{1}x_{3} \wedge \forall x_{2}\neg Fx_{2}y_{1}y_{1}) & \label{embeisp}
\end{eqnarray}

$x_{1}$ must be replaced with $y_{1}$ to unify $\{Fy_{1}x_{1}x_{3}, \neg Fx_{2}y_{1}y_{1}\}$.
However, $\forall$E cannot be applied to (\ref{embeisp}) to achieve this because $\exists y_{1}$ is in the scope of $\forall x_{1}$.
Independent of $\exists$M-optimization, one must apply $\wedge$I to replace a derivate of $x_{1}$ with a derivate of $y_{1}$; cf. Definition \ref{derivate}.
Applying $\wedge$I once to multiply $\forall x_{1}$ results in the following formula:
\begin{subequations}
\begin{eqnarray}
& \forall x_{1_{1}} \exists y_{1_{1}}(\forall x_{3_{1}} Fy_{1_{1}}x_{1_{1}}x_{3_{1}} \wedge \forall x_{2_{1}}\neg Fx_{2_{1}}y_{1_{1}}y_{1_{1}}) \; \wedge & \label{embeisp11}\\
& \forall x_{1_{2}} \exists y_{1_{2}}(\forall x_{3_{2}} Fy_{1_{2}}x_{1_{2}}x_{3_{2}} \wedge \forall x_{2_{2}}\neg Fx_{2_{2}}y_{1_{2}}y_{1_{2}}) & \label{embeisp12}
\end{eqnarray}
\end{subequations}
This formula allows one to derive an explicit contradiction by unifying the pair $\{Fy_{1_{1}}x_{1_{1}}x_{3_{1}},$ $\neg Fx_{2_{2}}y_{1_{2}}y_{1_{2}}\}$ through the application of $\forall$E to a corresponding optimal prenex normal form.
Since unifying the stated pair of literals is sufficient to derive an explicit contradiction, we delete the other two literals:

\begin{eqnarray}
& \forall x_{1_{1}} \exists y_{1_{1}} \forall x_{3_{1}} Fy_{1_{1}}x_{1_{1}}x_{3_{1}} \wedge \exists y_{1_{2}}\forall x_{2_{2}}\neg Fx_{2_{2}}y_{1_{2}}y_{1_{2}} & \label{embeisp2}
\end{eqnarray}

Applying PN laws allows one to generate the following optimal prenex normal form from (\ref{embeisp2}):
\begin{eqnarray}
& \exists y_{1_{2}}\forall x_{1_{1}}\exists y_{1_{1}}\forall x_{3_{1}}\forall x_{2_{2}}(Fy_{1_{1}}x_{1_{1}}x_{3_{1}} \wedge \neg Fx_{2_{2}}y_{1_{2}}y_{1_{2}}) & \label{embeisp3}
\end{eqnarray}
This makes it possible to apply $\forall$E such that the pair of literals in (\ref{embeisp3})  is unified by the following substitutions:
\begin{eqnarray}
& \{\{x_{1_{1}},y_{1_{2}}\},\{x_{2_{2}},y_{1_{1}}\},\{x_{3_{1}},y_{1_{2}}\}\}. &
\end{eqnarray}

However, the application of $\wedge$I that is involved in this proof is apparently only a cumbersome surrogate for a corresponding $\exists$M-optimization. This is so because $\exists$M-optimization
directly leads to the following formula, which is identical to (\ref{embeisp2}) despite the naming of the $x$ variables:
\begin{eqnarray}
& \forall x_{1} \exists y_{1_{1}} \forall x_{3} Fy_{1_{1}}x_{1}x_{3} \wedge \exists y_{1_{2}}\forall x_{2}\neg Fx_{2}y_{1_{2}}y_{1_{2}} & \label{embeisp1}
\end{eqnarray}
\end{exa}
%\vspace{0.25cm}

\begin{thm}\label{noandI}
$\exists$M-optimization makes the application of $\wedge$I superfluous in
proofs of contradiction for subformulae $\psi$.
\end{thm}

\proof
Applications of $\wedge$I are necessary for unification as long as one does not refer to %anti-prenex formulae $\psi'$ and
$\exists$M-optimized formulae $\psi''$ in the following cases:
\begin{enumerate} \label{argemstattandI}
\item to replace a universal variable $\nu$ with an existential variable $\mu$, where $\exists \mu$ is in the scope of $\forall \nu$ in $\psi$
(case 1; cf.  `suboptimal prenex', p. \pageref{suboptimalprae}), or
\item to separate a universal variable $\nu$ in $\psi$ such that the different variables $\nu_{1}$ and $\nu_{2}$ can be replaced with different $y$ variables
(case 2; cf. `$x$ ambiguity', p. \pageref{xueindeutigkeit}).
\end{enumerate}
However, these substitutions can be achieved through $\exists$M-optimization as long as the substitutions are logically valid.
The conditions for case 1 or case 2 only hold in $\exists$M-optimized expressions $\psi''$ if $\exists \rho$ is in the scope of $\forall \nu$ and above $\wedge$ and
if multiplying $\exists \mu$ is incompatible with unifying the pair of literals in $\psi'$. Any application of $\wedge$I to $\forall \nu$ would necessarily also multiply $\exists \rho$.
Therefore, applications of $\wedge$I cannot enable unifications that cannot be achieved through $\exists$M-optimization.
\qed

\begin{exa}
The following formulae exemplify the two cases described above:
\begin{eqnarray}
& \forall x_{1} \exists y_{1} (\exists y_{2} Fy_{1}y_{2} \wedge \neg Fy_{1}x_{1})  & \label{B3B}\\
& \exists y_{1}\exists y_{2}\forall x_{1} \exists y_{3} (\exists y_{2} Fy_{3}y_{1}x_{1} \wedge \neg Fy_{3}x_{1}y_{2}) & \label{B2B}
\end{eqnarray}
In (\ref{B3B}), $\nu = x_{1}$, $\rho = y_{1}$, and $\mu = y_{2}$; in (\ref{B2B}),  $\nu = x_{1}$, $\mu_{1} = y_{1}$, $\mu_{2} = y_{2}$, and $\rho = y_{3}$.
To replace $x_{1}$ in (\ref{B3B}) with $y_{1}$,  or to replace $x_{1}$ in L1 of (\ref{B2B}) with $x_{1_{1}}$ and in L2 of (\ref{B2B}) with $x_{1_{2}}$ such that
$x_{1_{1}}$ can be replaced with $y_{2}$ and $x_{1_{2}}$ with $y_{1}$, one must multiply $\exists y_{1}$ in (\ref{B3B}) and $\exists y_{3}$ in (\ref{B2B}).
This applies to both $\exists$M-optimization and the application of $\wedge$I. However, in both cases, this is incompatible with unifying the first position of both literals.
\end{exa}

\begin{thm}\label{undIeinmalth}
One can generate $\psi''$ from $\psi$ without applying $\exists$M by applying $\wedge$I once.
\end{thm}

\proof\label{undIeinmal}
Let $\exists \mu$ be the outermost existential quantifier that is multiplied in Algorithm \ref{mem}.
Then, one can obtain $\psi''$ from $\psi$ by applying the following algorithm:
\begin{enumerate}
\item Apply Algorithm \ref{malg} to minimize scopes and rename variables.
\item Multiply $\exists \mu A(\mu)$ by applying $\wedge$I once.
\item Delete L2 from the first conjunct along with all quantifiers binding variables that occur only in L2.
\item Delete L1 from the second conjunct along with all quantifiers binding variables that occur only in L1.
\item Apply Algorithm \ref{malg} to minimize scopes and rename variables.
\end{enumerate}
\qed

Thus, in a $\wedge$I-minimal proof strategy for subformulae $\psi$ in the NNF-calculus without $\exists$M, $\wedge$I will be applied at most once.
%In calling $\psi''$ an anti-prenex, $\exists$M-optimized and with $\psi$ sat-equivalent formula it does not matter whether it is generated by applying $\exists$M or by applying the algorithm described in the proof of Theorem \ref{undIeinmalth} that includes $\wedge$I.

To decide whether $\psi$ is contradictory, it is sufficient to transform $\psi$ into a sat-equivalent, $\exists$M-optimized formula $\psi''$; cf. section \ref{subkpairs}.
In the following, I will first describe how to compute the substitutions that are necessary for unifying the pairs of literals in $\psi''$ (section \ref{substis}). Afterward, I will describe how to generate all optimized prenex normal forms $\psi'''$ from $\psi''$ (section \ref{wpopt}).

\section{Substitution List $\sigma$}\label{substis}

To decide whether $\psi$ is contradictory, a substitution list $\sigma$ is generated. For each $x$ variable, the list specifies the $y$ variables with which it
must be replaced to unify the pair of connected literals $\{L1,L2\}$ in $\psi''$. Such a list does not imply that $\{L1,L2\}$ is unifiable. It may well be that some $x$ variables
would need to be replaced with several $y$ variables to unify $\{L1,L2\}$.
The following algorithm generates $\sigma$ for a given pair of literals $\{L1,L2\}$:

\newpage

\begin{algo}\hfill \label{sigma}
\begin{enumerate}
\item Generate a list $\sigma1$ of all pairs of variables $\{v_{1},v_{2}\}$ with $v_{1}$ in position $n$ in L1 and $v_{2}$ in the same position $n$ in L2,
where $v_{1}$ or $v_{2}$ (or both) is an $x$ variable.
\item To each level-1 list that contains an $x$ variable $\nu_{1}$ and exactly one $y$ variable $\mu$, recursively add those $x$ variables $\nu_{2}$ ($\nu_{1} \neq \nu_{2}$) that
occur in some other level-1 list together with $\nu_{1}$. Delete any level-1 list that contains only $x$ variables if all of those $x$ variables also occur in some level-1 list that contains a $y$ variable. Repeat this procedure until no further changes occur. The resulting list is denoted by $\sigma2$.
\item Generate a list $\sigma3$ consisting of level-1 lists that each contain one $x$ variable $\nu$ and all $y$ variables that occur together with $\nu$ in some level-1 list in $\sigma2$. To generate the level-1 lists in $\sigma3$, traverse all $x$ variables $\nu$ that occur in $\sigma2$ and take all $y$ variables that are members of lists in $\sigma2$ of which $\nu$ is also a member. If $\nu$ is not a member of any list in $\sigma2$ that contains a $y$ variable, add $\{\nu,y_{0}\}$ to $\sigma3$.
    Per its definition, $y_{0}$ does not occur in $\psi$, $\psi'$ or $\psi''$. The resulting list is the substitution list $\sigma$ that specifies how each $x$ variable is to be substituted such that $\{L1,L2\}$ is unified.
\end{enumerate}
\end{algo}

\begin{exa}\label{litb} Let $\{L1,L2\}$ be the following pair:

\begin{eqnarray}
& \{Fx_{1}x_{2}x_{2}x_{4}, \neg Fy_{2}x_{1}x_{3}x_{4}\} & \label{litbeisp}
\end{eqnarray}
\begin{description}
\item[$\sigma$1] $\{\{x_{1},y_{2}\}, \{x_{2},x_{1}\}, \{x_{2},x_{3}\}, \{x_{4},x_{4}\}\}$.
\item[$\sigma$2] $\{\{x_{1},x_{2},x_{3},y_{2}\}, \{x_{4},x_{4}\}\}$.
\item[$\sigma$3 (=$\sigma(\ref{litbeisp})$)] $\{\{x_{1},y_{2}\}, \{x_{2},y_{2}\}, \{x_{3},y_{2}\},\{x_{4},y_{0}\}\}$.
\end{description}
\end{exa}

\begin{defi}
\emph{$x$ lists} are the members (= level-1 lists) of the substitution list $\sigma$.
\end{defi}

\begin{defi}
An $x$ list is \emph{unambiguous} iff it contains exactly one $y$ variable.
$\sigma$ is \emph{unambiguous} iff all $x$ lists in $\sigma$ are unambiguous.
\end{defi}

\begin{exa}
The $x$ lists of $\sigma$3 in Example \ref{litb} and, therefore, $\sigma$(\ref{litbeisp}) \label{separierung}
are unambiguous.
\end{exa}

\begin{defi}
A substitution list $\sigma$ is \emph{maximally separated} iff the number of members in $\sigma$ cannot be increased any further by separating $x$ variables through equivalence transformations.
\end{defi}

Since Algorithm \ref{mem} minimizes the scopes of universal quantifiers to the maximal extent under equivalence transformation and, thus, increases the number of
universal quantifiers to the maximal extent (cf. Theorem \ref{minmax}), the substitution list of $\psi''$ is maximally separated.
Unlike the substitution list of $\psi$, the substitution list of $\psi''$ is necessarily maximally separated but is not necessarily unambiguous.

\begin{exa}
Let $\psi''$ be as follows:
\begin{eqnarray}
& \exists y_{1} \exists y_{2} \forall x_{1} \exists y_{3}(Fy_{1}x_{1}y_{3} \wedge \neg Fx_{1}y_{2}y_{3}) & \label{maxsep}
\end{eqnarray}
The corresponding substitution list is $\sigma$(\ref{maxsep}) = $\{\{x_{1},y_{1},y_{2}\}\}$. This substitution list is not unambiguous.
However, $\exists$M cannot be applied to (\ref{maxsep}) because the conditions for the direct case are satisfied.
Hence, $\sigma(\ref{maxsep})$ is maximally separated but not unambiguous.
\end{exa}

A decision on the unifiability of a pair of connected literals refers to a \emph{maximally separated} substitution list.
A %proof by contradiction
proof of contradiction according to the $\wedge$I-minimal proof strategy in the NNF-calculus, however, requires an \emph{unambiguous} substitution list.

Algorithm \ref{sigma} generates not \emph{minimal} but \emph{maximal} \label{maxi} substitution lists. In $\{Fx_{1}x_{1}y_{1},$ $\neg Fx_{2}y_{2}x_{2}\}$, for example,
$\sigma = \{\{x_{1},y_{1},y_{2}\}, \{x_{2},y_{1},y_{2}\}\}$: both $x$ lists contain both $y_{1}$ and $y_{2}$, which is not necessary.
It is only necessary that $x_{1}$ be substituted with $y_{2}$ in position 2 of L1, that $x_{2}$ be substituted with $y_{1}$ in position 3 of L2, and that $x_{1}$ and $x_{2}$ be substituted with one and the same
$y$ variable in position 1 of both L1 and L2. This last condition is compatible with two different \emph{minimal} substitution lists: $\{\{x_{1},y_{1},y_{2}\}, \{x_{2},y_{1}\}\}$ or, alternatively,
$\{\{x_{1},y_{2}\}, \{x_{2},y_{1},y_{2}\}\}$. However, when deciding whether subformulae $\psi$ are contradictory, one can abstain from generating alternative minimal substitution lists.
It is sufficient to refer to a maximal substitution list $\sigma$ because -- as proven in section \ref{bkrit} -- a necessary condition for the contradiction of a subformula $\psi$ is that the $\sigma$ of $\psi''$ be unambiguous, which implies that there is only one minimal substitution list in this case. If the maximal substitution list $\sigma$ is not unambiguous, this implies that \emph{at least one} $x$ variable must be replaced with more than one $y$ variable to unify the pair of literals in $\psi''$. This is sufficient to conclude that $\psi''$ and, therefore, $\psi$ are not contradictory; cf. Theorem \ref{b1b2}, condition C1, p. \pageref{b1b2}.

Section \ref{subkpairs} will explain how it can be decided whether the pair of literals contained in $\psi''$ is unifiable by referring to the
syntactic properties of $\psi''$ (without any further equivalence transformation). In the following section, however, I will describe a costlier algorithm for such decision-making.
This algorithm generates the set of all optimized prenex normal forms $\psi'''$ from the anti-prenex, $\exists$M-optimized normal form $\psi''$.
I present this algorithm first because it serves as the foundation for the syntactic criteria defined in section \ref{subkpairs} that directly refer to $\psi''$.

\section{Optimized Prenexes}\label{wpopt}

This section describes an algorithm that generates all optimized prenex normal forms (cf. Definition \ref{optimiert}) from an anti-prenex, $\exists$M-optimized and rectified formula $\psi''$.
%The algorithm considers all alternatives to place existential quantifiers left to universal quantifiers if possible by applying PN-laws.
Since $\psi''$ does not contain $\vee$, only PN1, PN2, PN5 and PN6  need to be considered.
%Der Algorithmus könnte effizienter gestaltet
%werden, indem nur in dem Fall optimierte Pränexe gebildet werden, in dem alle $x$-Listen aus $\sigma$ eindeutig sind (s.u. Theorem \ref{b1b2}, Bedingung B1) und die Bildung eines optimierten
%Pränex abgebrochen wird, sobald ein $\exists \mu$ im Wirkungsbereich eines
%Allquantors $\forall \nu$ steht und $\{\nu,\mu\} \in \sigma$ (s.u. Theorem \ref{b1b2}, Bedingung B2). Hingegen kommt es uns an dieser
%Stelle nicht auf einen effizienten Algorithmus an, sondern auf einen Algorithmus, der die logischen Grundlagen
%erklärt. Wie in Abschnitt \ref{subkpairs} erläutert wird, kann für die Identifikation unifizierbarer Literalpaare die umständliche Prozedur der Bildung von $\psi''$ weitgehend und die der
%Bildung von $\psi'''$ vollends umgangen werden.
%
%Der an dieser Stelle entscheidende, grundlegende Punkt ist, dass
%nur Bildung $\exists$M-optimierter, anti-pränexer Normalformen $\psi''$ überhaupt die Bildung optimierter pränexer
%Normalformen ermöglicht, in denen Allquantoren möglichst nach innen gezogen werden. \label{praefix}

To generate the set of all optimized prenex normal forms $\psi'''$ from $\psi''$ by applying PN laws from right to left, proceed as follows:

\vspace{0.25cm}

\begin{algo}\hfill \label{wpA}
\begin{enumerate}
\item $\psi^{+} = \{\psi''\}$.
\item $\psi''' = \{\}$.
\item If $\psi^{+} \neq \{\}$, select the first member $\psi^{+}_{1}$ from $\psi^{+}$ and go to (a); otherwise, return $\psi'''$ and terminate.
\begin{enumerate}
\item Let $\psi^{+}_{1}$ be of the form $Q$$(A \wedge B)$, where $Q$ is a sequence of quantifiers of length $m$ ($m \geq 0$).
If neither $A$ nor $B$ contains any quantifier, add $\psi^{+}_{1}$ to $\psi'''$, delete $\psi^{+}_{1}$ in $\psi^{+}$ and go to (3); otherwise, go to (b).
\item If PN6 can be applied to $A \wedge B$, then $\psi^{+}_{1}$ = $Q$PN6$[A \wedge B]$ and go to (a); otherwise, go to (c).
\item If PN5 can be applied to $A \wedge B$, then $\psi^{+}_{1}$ = $Q$PN5$[A \wedge B]$ and go to (a); otherwise, go to  (d).
\item If $A$ contains only universal quantifiers, then $\psi^{+}_{1}$ = $Q$PN1$[A \wedge B]$ and go to (a); otherwise, go to  (e).
\item If $B$ contains only universal quantifiers, then $\psi^{+}_{1}$ = $Q$PN2$[A \wedge B]$ and go to (a); otherwise, go to  (f).
\item Delete $\psi^{+}_{1}$ in $\psi^{+}$ and add $Q$PN2$[A \wedge B]$ and $Q$PN1$[A \wedge B$] to $\psi^{+}$. Go to (3).
\end{enumerate}
\end{enumerate}
\end{algo}

\begin{exa} Let $\psi''$ be as follows:

\begin{eqnarray}
& \exists y_{1}\forall x_{1}\exists y_{2}\forall x_{2} Fy_{1}x_{1}y_{2}x_{2} \wedge \forall x_{3}\exists y_{3}\forall x_{4}\forall x_{5} \neg Fx_{3}y_{3}x_{4}x_{5} & \label{prae}
\end{eqnarray}

The following schema illustrates the application of Algorithm (\ref{wpA}) to generate $\psi^{+}$ and $\psi'''$ from (\ref{prae}).

\renewcommand{\arraystretch}{1.5}
\begin{longtable}{|c|c|c|}\hline
{\bf No.} & {\bf $\psi^{+}; \psi'''$} & {\bf Rule}\\\hline
(1) &  \begin{tabular}{ll}$\psi^{+} = $ & $\{\exists y_{1}\forall x_{1}\exists y_{2}\forall x_{2} Fy_{1}x_{1}y_{2}x_{2} \wedge \forall x_{3}\exists y_{3}\forall x_{4}\forall x_{5} \neg Fx_{3}y_{3}x_{4}x_{5}\}$;\hspace{0.25cm}\\
$\psi''' =$ & $\{\}$
\end{tabular} & 1, 2 \\\hline
(2) & \hspace{-0.2cm} \begin{tabular}{ll}$\psi^{+} = $ & $ \{\exists y_{1}(\forall x_{1}\exists y_{2}\forall x_{2} Fy_{1}x_{1}y_{2}x_{2} \wedge \forall x_{3}\exists y_{3}\forall x_{4}\forall x_{5} \neg Fx_{3}y_{3}x_{4}x_{5})\}$;\\
$\psi'''$ & $\{\}$ \end{tabular} &  3(b) \\ \hline
(3) & \begin{tabular}{ll}$\psi^{+} = $ & \begin{tabular}{l}$\{\exists y_{1}\forall x_{1}(\exists y_{2}\forall x_{2} Fy_{1}x_{1}y_{2}x_{2} \wedge \forall x_{3}\exists y_{3}\forall x_{4}\forall x_{5} \neg Fx_{3}y_{3}x_{4}x_{5})$,\\
$\exists y_{1}\forall x_{3}(\forall x_{1}\exists y_{2}\forall x_{2} Fy_{1}x_{1}y_{2}x_{2} \wedge \exists y_{3}\forall x_{4}\forall x_{5} \neg Fx_{3}y_{3}x_{4}x_{5})\}$\end{tabular};\\
$\psi''' =$ & $\{\}$
\end{tabular} & 3(f) \\\hline
(4) &\begin{tabular}{ll}$\psi^{+} = $ &  \begin{tabular}{l}$\{\exists y_{1}\forall x_{1}\exists y_{2}(\forall x_{2} Fy_{1}x_{1}y_{2}x_{2} \wedge \forall x_{3}\exists y_{3}\forall x_{4}\forall x_{5} \neg Fx_{3}y_{3}x_{4}x_{5})$,\\
$\exists y_{1}\forall x_{3}(\forall x_{1}\exists y_{2}\forall x_{2} Fy_{1}x_{1}y_{2}x_{2} \wedge \exists y_{3}\forall x_{4}\forall x_{5} \neg Fx_{3}y_{3}x_{4}x_{5})\}$\end{tabular};\\
$\psi''' =$ & $\{\}$\end{tabular} & 3(b) \\\hline
(5) &\begin{tabular}{ll}$\psi^{+} = $ &  \begin{tabular}{l}$\{\exists y_{1}\forall x_{1}\exists y_{2}\forall x_{3}(\forall x_{2} Fy_{1}x_{1}y_{2}x_{2} \wedge \exists y_{3}\forall x_{4}\forall x_{5} \neg Fx_{3}y_{3}x_{4}x_{5})$,\\
$\exists y_{1}\forall x_{3}(\forall x_{1}\exists y_{2}\forall x_{2} Fy_{1}x_{1}y_{2}x_{2} \wedge \exists y_{3}\forall x_{4}\forall x_{5} \neg Fx_{3}y_{3}x_{4}x_{5})\}$\end{tabular};\\
$\psi''' =$ & $\{\}$\end{tabular} & 3(d) \\ \hline
(6) & \begin{tabular}{ll}$\psi^{+} = $ &  \begin{tabular}{l}$\{\exists y_{1}\forall x_{1}\exists y_{2}\forall x_{3}\exists y_{3}(\forall x_{2} Fy_{1}x_{1}y_{2}x_{2} \wedge \forall x_{4}\forall x_{5} \neg Fx_{3}y_{3}x_{4}x_{5})$,\\
$\exists y_{1}\forall x_{3}(\forall x_{1}\exists y_{2}\forall x_{2} Fy_{1}x_{1}y_{2}x_{2} \wedge \exists y_{3}\forall x_{4}\forall x_{5} \neg Fx_{3}y_{3}x_{4}x_{5})\}$\end{tabular};\\
$\psi''' =$ & $\{\}$\end{tabular} & 3(c) \\ \hline
(7) & \begin{tabular}{ll}$\psi^{+} = $ &  \begin{tabular}{l}$\{\exists y_{1}\forall x_{1}\exists y_{2}\forall x_{3}\exists y_{3}\forall x_{4}(\forall x_{2} Fy_{1}x_{1}y_{2}x_{2} \wedge \forall x_{5} \neg Fx_{3}y_{3}x_{4}x_{5})$,\\
$\exists y_{1}\forall x_{3}(\forall x_{1}\exists y_{2}\forall x_{2} Fy_{1}x_{1}y_{2}x_{2} \wedge \exists y_{3}\forall x_{4}\forall x_{5} \neg Fx_{3}y_{3}x_{4}x_{5})\}$\end{tabular};\\
$\psi''' =$ & $\{\}$\end{tabular} & 3(d) \\ \hline
(8) & \begin{tabular}{ll}$\psi^{+} = $ &  \begin{tabular}{l}$\{\exists y_{1}\forall x_{1}\exists y_{2}\forall x_{3}\exists y_{3}\forall x_{4}\forall x_{5}(\forall x_{2} Fy_{1}x_{1}y_{2}x_{2} \wedge \neg Fx_{3}y_{3}x_{4}x_{5})$,\\
$\exists y_{1}\forall x_{3}(\forall x_{1}\exists y_{2}\forall x_{2} Fy_{1}x_{1}y_{2}x_{2} \wedge \exists y_{3}\forall x_{4}\forall x_{5} \neg Fx_{3}y_{3}x_{4}x_{5})\}$\end{tabular};\\
$\psi''' =$ & $\{\}$\end{tabular} & 3(d) \\ \hline
(9) & \begin{tabular}{ll}$\psi^{+} = $ &  \begin{tabular}{l}$\{\exists y_{1}\forall x_{1}\exists y_{2}\forall x_{3}\exists y_{3}\forall x_{4}\forall x_{5}\forall x_{2}( Fy_{1}x_{1}y_{2}x_{2} \wedge \neg Fx_{3}y_{3}x_{4}x_{5})$,\\
$\exists y_{1}\forall x_{3}(\forall x_{1}\exists y_{2}\forall x_{2} Fy_{1}x_{1}y_{2}x_{2} \wedge \exists y_{3}\forall x_{4}\forall x_{5} \neg Fx_{3}y_{3}x_{4}x_{5})\}$\end{tabular};\\
$\psi''' =$ & $\{\}$\end{tabular}& 3(e) \\ \hline
(10) & \begin{tabular}{ll}$\psi^{+} =$ & $\{\exists y_{1}\forall x_{3}(\forall x_{1}\exists y_{2}\forall x_{2} Fy_{1}x_{1}y_{2}x_{2} \wedge \exists y_{3}\forall x_{4}\forall x_{5} \neg Fx_{3}y_{3}x_{4}x_{5})\};$ \\
$\psi''' = $ & $\{\exists y_{1}\forall x_{1}\exists y_{2}\forall x_{3}\exists y_{3}\forall x_{4}\forall x_{5}\forall x_{2}(Fy_{1}x_{1}y_{2}x_{2} \wedge \neg Fx_{3}y_{3}x_{4}x_{5})\}$
\end{tabular} & 3(a) \\ \hline
(11) & \begin{tabular}{ll}$\psi^{+} =$ & $\{\exists y_{1}\forall x_{3}\exists y_{3}(\forall x_{1}\exists y_{2}\forall x_{2} Fy_{1}x_{1}y_{2}x_{2} \wedge \forall x_{4}\forall x_{5} \neg Fx_{3}y_{3}x_{4}x_{5})\};$ \\
$\psi''' = $ & $\{\exists y_{1}\forall x_{1}\exists y_{2}\forall x_{3}\exists y_{3}\forall x_{4}\forall x_{5}\forall x_{2}(Fy_{1}x_{1}y_{2}x_{2} \wedge \neg Fx_{3}y_{3}x_{4}x_{5})\}$
\end{tabular} & 3(c) \\ \hline
(12) & \begin{tabular}{ll}$\psi^{+} =$ & $\{\exists y_{1}\forall x_{3}\exists y_{3}\forall x_{1}(\exists y_{2}\forall x_{2} Fy_{1}x_{1}y_{2}x_{2} \wedge \forall x_{4}\forall x_{5} \neg Fx_{3}y_{3}x_{4}x_{5})\};$ \\
$\psi''' = $ & $\{\exists y_{1}\forall x_{1}\exists y_{2}\forall x_{3}\exists y_{3}\forall x_{4}\forall x_{5}\forall x_{2}(Fy_{1}x_{1}y_{2}x_{2} \wedge \neg Fx_{3}y_{3}x_{4}x_{5})\}$
\end{tabular} & 3(e) \\ \hline
(13) & \begin{tabular}{ll}$\psi^{+} =$ & $\{\exists y_{1}\forall x_{3}\exists y_{3}\forall x_{1}\exists y_{2}(\forall x_{2} Fy_{1}x_{1}y_{2}x_{2} \wedge \forall x_{4}\forall x_{5} \neg Fx_{3}y_{3}x_{4}x_{5})\};$ \\
$\psi''' = $ & $\{\exists y_{1}\forall x_{1}\exists y_{2}\forall x_{3}\exists y_{3}\forall x_{4}\forall x_{5}\forall x_{2}(Fy_{1}x_{1}y_{2}x_{2} \wedge \neg Fx_{3}y_{3}x_{4}x_{5})\}$
\end{tabular} & 3(b) \\ \hline
(14) & \begin{tabular}{ll}$\psi^{+} =$ & $\{\exists y_{1}\forall x_{3}\exists y_{3}\forall x_{1}\exists y_{2}\forall x_{4}(\forall x_{2} Fy_{1}x_{1}y_{2}x_{2} \wedge \forall x_{5} \neg Fx_{3}y_{3}x_{4}x_{5})\};$ \\
$\psi''' = $ & $\{\exists y_{1}\forall x_{1}\exists y_{2}\forall x_{3}\exists y_{3}\forall x_{4}\forall x_{5}\forall x_{2}(Fy_{1}x_{1}y_{2}x_{2} \wedge \neg Fx_{3}y_{3}x_{4}x_{5})\}$
\end{tabular} & 3(d) \\ \hline
(15) & \begin{tabular}{ll}$\psi^{+} =$ & $\{\exists y_{1}\forall x_{3}\exists y_{3}\forall x_{1}\exists y_{2}\forall x_{4}\forall x_{5} (\forall x_{2} Fy_{1}x_{1}y_{2}x_{2} \wedge \neg Fx_{3}y_{3}x_{4}x_{5})\};$ \\
$\psi''' = $ & $\{\exists y_{1}\forall x_{1}\exists y_{2}\forall x_{3}\exists y_{3}\forall x_{4}\forall x_{5}\forall x_{2}(Fy_{1}x_{1}y_{2}x_{2} \wedge \neg Fx_{3}y_{3}x_{4}x_{5})\}$
\end{tabular} & 3(d) \\ \hline
(16) & \begin{tabular}{ll}$\psi^{+} =$ & $\{\exists y_{1}\forall x_{3}\exists y_{3}\forall x_{1}\exists y_{2}\forall x_{4}\forall x_{5}\forall x_{2}(Fy_{1}x_{1}y_{2}x_{2} \wedge \neg Fx_{3}y_{3}x_{4}x_{5})\};$ \\
$\psi''' = $ & $\{\exists y_{1}\forall x_{1}\exists y_{2}\forall x_{3}\exists y_{3}\forall x_{4}\forall x_{5}\forall x_{2}(Fy_{1}x_{1}y_{2}x_{2} \wedge \neg Fx_{3}y_{3}x_{4}x_{5})\}$
\end{tabular} & 3(e) \\ \hline
(17) & \begin{tabular}{ll}$\psi^{+} =$ & $\{\};$ \\ $\psi''' = $ & \begin{tabular}{ll}
$\{\exists y_{1}\forall x_{1}\exists y_{2}\forall x_{3}\exists y_{3}\forall x_{4}\forall x_{5}\forall x_{2}(Fy_{1}x_{1}y_{2}x_{2} \wedge \neg Fx_{3}y_{3}x_{4}x_{5})$,\\
$\exists y_{1}\forall x_{3}\exists y_{3}\forall x_{1}\exists y_{2}\forall x_{4}\forall x_{5}\forall x_{2}(Fy_{1}x_{1}y_{2}x_{2} \wedge \neg Fx_{3}y_{3}x_{4}x_{5})\}$
\end{tabular}\end{tabular} & 3(a) \\\hline
\end{longtable}
\end{exa}

\begin{thm}\label{wpaequi}
Algorithm \ref{wpA} generates only prenex normal forms that are logically equivalent to $\psi''$.
\end{thm}

\proof
This theorem follows from the fact that the PN laws are equivalence rules that are applied from right to left in Algorithm \ref{wpA}.
\qed

\begin{thm}\label{wpaterm}
Algorithm \ref{wpA} terminates.
\end{thm}

\proof
This theorem follows from the fact that only a finite number of quantifiers can exist below $\wedge$ in the logical hierarchy of $\psi''$.
\qed

\begin{thm}\label{alleprae}
Algorithm \ref{wpA} generates the set of all possible optimized prenex normal forms $\psi'''$ from $\psi''$.
\end{thm}

\proof
This theorem follows from the facts that in Algorithm \ref{wpA}, PN5 and PN6 are applied prior to PN1 and PN2, and two alternative prenex forms are generated in the case of 3(g).
\qed

\begin{defi}\label{optimal}
An \emph{optimal} prenex normal form $\psi^{*}$ of $\psi''$ is a prenex normal form in $\psi'''$ such that for all $x$ lists $\{\nu, \ldots \mu_{j} \ldots\}$ ($j>0$),
$\forall \nu$ is to the right of each existential quantifier $\exists \mu_{j}$.
\end{defi}

\begin{exa}
\begin{description}
\item[$\sigma(\ref{prae})$] $\{\{x_{1},y_{3}\},\{x_{2},y_{0}\},\{x_{3},y_{1}\},\{x_{4},y_{2}\},\{x_{5},y_{0}\}\}.$
\end{description}
Hence, only the second prenex normal form in $\psi'''$ from line (17) in the schema above is optimal.
\end{exa}

\section{sat-Criteria for $\psi$}\label{bkrit}

\begin{thm}\label{b1b2}
$\psi$ is not contradictory iff one of the two following conditions is satisfied:
\begin{description}
\item[C1] The maximally separated substitution list $\sigma$ of the pair of connected literals in $\psi''$ (and, consequently, of each formula from $\psi'''$) is ambiguous.
%\begin{description}
%\item[B1A:] $\sigma$ ein Element der Form $\{y_{i},y_{j}\}$ enthält, wobei $i\neq j$ (= $y$-Unein\-deu\-tig\-keit), oder \label{B1A}
%$\sigma$ eine $x$-Liste mit einer $x$-Variable und mehr als einer $y$-Variable enthält (= $x$-Uneindeutigkeit). \label{B1B}
\item[C2] No optimal prenex normal form $\psi^{*}$ exists in $\psi'''$.\label{B2}
\end{description}
\end{thm}

C1 concerns the substitution list $\sigma$ of $\psi''$, whereas C2 concerns the optimized prenexes in $\psi'''$. These are the relevant features on which the unifiability of the pair of connected literals in $\psi''$ depends.
The unifiability of a pair of literals can be decided by generating $\psi$, $\psi'$, $\psi''$, $\psi'''$, and $\sigma$ to determine whether C1 or C2 is satisfied.

\proof
We prove that Theorem \ref{b1b2} holds in the direction from left to right
by proving the contrapositive: `If neither C1 nor C2 is satisfied, then $\psi$ is contradictory'.
This follows from the following two points:
 \begin{enumerate}
 \item If C2 is not satisfied, then some optimal prenex form $\psi^{*}$ exists, and if C1 is not satisfied, then some unambiguous substitution list $\sigma$ exists.
 Under such conditions, the pair of connected literals in $\psi^{*}$ can be unified by applying $\forall$E to replace each $x$ variable in $\sigma$ with the $y$ variable in the corresponding $x$ list of $\sigma$.
 \item The transformation of $\psi$ into $\psi^{*}$ is a sat-equivalence transformation according to Theorems \ref{memsat} and \ref{wpaequi}.
  \end{enumerate}
\qed

%Wenn hingegen keine der Bedingungen B1 und B2 erfüllt sind, dann ist $\sigma$ nicht nur eine maximal separierte, sondern
%eine eindeutige Substitutionsliste und unter den optimierten pränexen Normalformen gibt es
%mindestens eine optimale pränexe Normalform $\psi^{*}$. Falls $x$-Variablen durch $y_{0}$ zu ersetzen sind, ist dies stets möglich, indem
%ein neuer Existenzquantor $\exists y_{0}$ eingeführt wird, der dem Pränex einer pränexen Normalform voransteht.
%Folglich kann mittels $\forall$E das Literalpaar $\{L1,L2\}$ aus $\psi$ unifiziert und ein expliziter Widerspruch abgeleitet werden.

In the proof of the right-to-left direction of Theorem \ref{b1b2}, C1 and C2 are reduced to `laws of invalidity' that specify syntactic properties
that prevent the application of $\forall$E for unification. The all-important point is that one can conclude from this that unification is impossible
if one refers to optimized prenex normal forms in $\psi'''$. The reason for this conclusion is that the conditions for replacing $x$ variables with $y$ variables when applying $\forall$E are optimal
in the case of the optimized prenex normal forms of $\psi'''$. Therefore, one can conclude that \emph{no} sat-equivalent prenex normal form exists that makes the substitutions that are necessary for unification according to $\sigma$ logically possible if no prenex normal form exists in $\psi'''$ that permits such a substitution.

%B1A beruht auf:
%
%\begin{eqnarray}
%& \ldots \exists \mu_{1} \ldots \exists \mu_{2} \ldots (L1(\mu_{1}) \wedge L2(\mu_{2})) \not \vdash \ldots \exists \mu \ldots (L1(\mu_{1}/\mu) \wedge L2(\mu_{2}/\mu)) & \; \mbox{B1A-R.} \nonumber
%\end{eqnarray}
%
%Laut Voraussetzung ist die Matrix der optimierten pränexen Normalformen in $\psi'''$
%eine Konjunktion $L1 \wedge L2$, wobei $\mu_{1}$ und $\mu_{2}$ an identischen
%Argumentstellen vorkommen. Unter dieser Voraussetzung ist eine Unifizierung ausgeschlossen.
%Auf B1-R. lässt sich z.B. (\ref{B1AR}) zurückführen.

\proof We prove that Theorem \ref{b1b2} holds in the direction from right to left: `If C1 or C2 is satisfied, then $\psi$ is not contradictory.'
According to Theorems \ref{memsat}, \ref{wpaequi} and \ref{alleprae}, $\psi'''$ contains all optimized prenex normal forms that are sat-equivalent to $\psi$.
It is necessary to show that C1 and C2 specify the conditions for $\psi$ to be non-contradictory.

If the maximally separated substitution list $\sigma$ is ambiguous, then at least one $x$ variable in $\psi''$ and in each formula in $\psi'''$ must be replaced with more than one $y$ variable. However, this is logically impossible. C1 is based on the following law of invalidity:
\begin{eqnarray}
\hspace{-0.55cm} & \ldots \exists \mu_{1} \ldots \exists \mu_{2} \ldots \forall \nu \varphi(\mu_{1},\mu_{2},\nu,\nu) \not \vdash \ldots \exists \mu_{1} \ldots \exists \mu_{2} \ldots
\varphi(\mu_{1},\mu_{2},\nu/\mu_{1},\nu/\mu_{2}) & \; \mbox{C1-R.} \nonumber
\end{eqnarray}
$\varphi(\mu_{1},\mu_{2},\nu,\nu)$ is an expression in which $\nu$ must be replaced with $\mu_{1}$ and also with $\mu_{2}$ to unify the pair of literals contained in
the expression. However, such a substitution is logically invalid because each application of $\forall$E can only replace $\nu$ with either $\mu_{1}$ or $\mu_{2}$
and, according to Theorems \ref{minmax} and \ref{noandI}, the number of universal quantifiers cannot be increased any further for the purpose of unification.
It is impossible to separate $\sigma$ any further through sat-equivalence transformations because $\sigma$ is already maximally separated. However, $\sigma$ would need to be further separated to make it possible to unify the pair of literals in question by applying $\forall$E.

C2 is based on the following law of invalidity for the set of optimized prenex normal forms $\psi'''$:
\begin{eqnarray}
& \ldots \forall \nu \ldots \exists \mu \varphi(\mu,\nu) \not \vdash \ldots \exists \mu \ldots \varphi(\mu, \nu/\mu) & \; \mbox{B2-R.} \nonumber
\end{eqnarray}
The application of $\forall$E requires that $\forall \nu$ be in the scope of $\exists \mu$.
According to Theorem \ref{alleprae}, Algorithm \ref{wpA} generates \emph{all} optimized prenex normal forms (= $\psi'''$). The quantifier
sequences of the prenex normal forms in $\psi'''$ cannot be optimized any further through sat-equivalence transformations, according to Theorems \ref{minmax} and \ref{noandI}.
If no optimized prenex normal form in $\psi'''$ permits the application of $\forall$E to achieve all substitutions that are necessary according to $\sigma$, then no optimal prenex normal form $\psi^{*}$ exists that enables %a proof by contradiction
one to prove that $\psi$ is contradictory.\qed

\begin{exa}
The following paradigmatic examples of anti-prenex, $\exists$M-optimized formulae $\psi''$ cannot be transformed into explicit contradictions because of condition C1:
\begin{eqnarray}
 & \exists y_{1}\exists y_{2} Fy_{1}y_{2} \wedge \forall x_{1} \neg Fx_{1}x_{1} \not \vdash \exists y_{1} \exists y_{2} (Fy_{1}y_{2} \wedge \neg Fy_{1}y_{2}) & \label{B1BFormel}\\
 & \hspace{-2cm}\exists y_{1}\forall x_{1}Fx_{1}x_{1}y_{1} \wedge \exists y_{2}\forall x_{2}\neg Fx_{2}y_{2}x_{2} \not \vdash \exists y_{1}\exists y_{2}(Fy_{2}y_{2}y_{1} \wedge \neg Fy_{2}y_{2}y_{1})  & \label{B1BFormel2}\\
&\hspace{-2cm} \exists y_{1} \exists y_{2} \forall x_{1} \exists y_{3}(Fy_{1}x_{1}y_{3} \wedge \neg Fx_{1}y_{2}y_{3}) \not\vdash
\exists y_{1} \exists y_{2} \exists y_{3}(Fy_{1}y_{2}y_{3} \wedge \neg Fy_{1}y_{2}y_{3}) &
\end{eqnarray}

Formula (\ref{B3B}) and the following paradigmatic examples of non-contradictory, anti-prenex, $\exists$M-optimized formulae $\psi''$  cannot be transformed into explicit contradictions because of condition C2:
\begin{eqnarray}
 & \forall x_{1}\exists y_{1} Fx_{1}y_{1} \wedge \forall x_{2} Fx_{2}x_{2} \not \vdash \exists y_{1}  (Fy_{1}y_{1}
\wedge \neg Fy_{1}y_{1}) & \label{B2AFormel}\\
& \forall x_{1} \exists y_{1}(Fx_{1}y_{1} \wedge \exists y_{2} \neg Fy_{2}y_{1}) \not \vdash \exists y_{1}\exists y_{2}(Fy_{2}y_{1} \wedge  \neg Fy_{2}y_{1}) & \\
& \forall x_{1} \exists y_{1} Fx_{1}y_{1} \wedge \forall x_{2}\exists y_{2} \neg Fy_{2}x_{2} \not \vdash \exists y_{1} \exists y_{2} (Fy_{2}y_{1} \wedge \neg Fy_{2}y_{1}) & \label{B2BFormel}
\end{eqnarray}
In each of these cases, no optimal prenex normal form $\psi^{*}$ can be generated from the anti-prenex, $\exists$M-optimized formula that would permit the necessary substitutions.
\end{exa}

\section{Optimized Search for Proofs of Contradiction}\label{optwid}

The decision procedure for subformulae $\psi$ can be used to conduct an optimized search for
proofs of contradiction for arbitrary NNFs $\phi$.
We presume that $\phi$ is transformed into an anti-prenex FOLDNF $\phi'$.\footnote{Further optimization is possible by transforming $\phi$ into a logically equivalent \emph{minimized}
FOLDNF; cf. \cite{Lampert}.} The single rectified disjuncts $D_{i}$ of $\phi'$ can be optimized by deleting literals in $D_{i}$ that are not members of any unifiable pair of literals from $D_{i}$; for details, cf. Algorithm \ref{bereinigung}.

To decide whether a pair of connected literals from $D_{i}$ is unifiable and thus capable of contributing to a
proof of contradiction, subformulae $\psi$ are generated.
Let $D_{i}$ be an NNF that contains at least two connected literals.
Subformulae $\psi$ with exactly two connected literals, L1 and L2, can then be generated from $D_{i}$ as follows:
\begin{algo}\hfill
\begin{enumerate}
\item Delete all literals in $\phi$ except L1 and L2.
\item Delete all quantifiers binding variables that do not occur in L1 or L2.
\item Delete all occurrences of $\wedge$ or $\vee$ except the occurrence that connects L1 and L2 in the logical hierarchy of $D_{i}$.
\item If L1 and L2 are connected by $\vee$, replace $\vee$ with $\wedge$.
\end{enumerate}
\end{algo}
Step (4) is justified by the fact that the unification of pairs of literals in
proofs of contradiction presumes that the literals are connected by conjunction.
This does not imply that the literals are also connected by conjunction in $D_{i}$. It may well be that
proofs of contradiction depend on the multiplication of universally quantified expressions by $\wedge$I to
replace the $x$ variables of the multiplied universal quantifier with different $y$ variables (including $y_{0}$). If the multiplied expression contains $\vee$, then literals that were connected by $\vee$ before the application of $\wedge$I are additionally connected by $\wedge$ after the application of $\wedge$I. A
proof of contradiction may then depend on the unification of those literals that are connected by $\wedge$ subsequent to the application of $\wedge$I.
Example \ref{bsporkpairbsp}, p. \pageref{bsporkpairbsp}, illustrates this: Subsequent to the application of $\wedge$I, occurrences of $Fx_{1}x_{2}x_{6}$ and $\neg Fx_{3}x_{1}x_{7}$ are connected not only by $\vee$ but also by $\wedge$. Consequently, the derivates
$Fx_{1_{2}}x_{2_{2}}x_{6_{2}}$ and $\neg Fx_{3_{1}}x_{1_{1}}x_{7_{1}}$ of $Fx_{1}x_{2}x_{6}$ and $\neg Fx_{3}x_{1}x_{7}$ are connected not by disjunction but by conjunction subsequent to maximal subscription, and this constitutes the proof of contradiction.

In general, a
proof of contradiction for a formula $\phi$ in accordance with the $\wedge$I-minimal proof strategy in the NNF-calculus is based on transforming $\phi$ into a sat-equivalent, optimized FOLDNF $\phi'$ and then specifying, for each disjunct $D_{i}$ in $\phi'$, (i) an anti-prenex, $\wedge$I-optimized disjunct $D_{i}^{*}$; (ii) a minimal set $\tt{L}$ of unifiable literals from $D_{i}^{*}$ that are unified in the
proof; (iii) a minimal substitution list $\sigma$ that specifies how each $x$ variable must be replaced with a $y$ variable to unify $\tt{L}$; and (iv) an optimal prenex that permits the substitutions necessary according to $\sigma$ and that can be generated from $D_{i}^{*}$.

Deciding whether a pair of connected literals in $D_{i}$ is unifiable is equivalent to deciding whether the corresponding subformula $\psi$ is contradictory. Since the latter is decidable by means of Theorem \ref{b1b2},
the identification of unifiable pairs of literals can be utilized to decide whether certain necessary conditions for a
proof of the refutability of
$D_{i}$ are satisfied.

\begin{thm}\label{dinonwid}
An NNF $\phi$ is not contradictory if the DNF matrix of a $D_{i}$ of the FOLDNF $\phi'$ that is sat-equivalent to $\phi$ contains a disjunct that does not contain any unifiable pair of literals.
\end{thm}

\proof
This theorem follows from the facts that $\phi$ is contradictory iff each disjunct $D_{i}$ of $\phi'$ is contradictory and that any $D_{i}$ is contradictory only if an explicit contradiction can be derived, which requires that each disjunct of the DNF matrix of $D_{i}$ must contain at least one unifiable pair of literals.
\qed

Thus, a necessary condition for contradiction can be specified with respect to sat-equivalence transformations to identify formulae that are not contradictory.

Independent of the generation of FOLDNFs and DNF matrices, the following can be stated:
\begin{thm}\label{herbrandth}
	$\wedge$-NNFs are decidable.
\end{thm}

\proof
According to Theorem \ref{subform}, a $\wedge$-NNF $\phi$ is contradictory iff at least one subformula $\psi$ from $\phi$ is contradictory and its refutability is decidable according to Theorem \ref{b1b2}.
\qed

\begin{exa}
The following formula is not contradictory:
\begin{eqnarray}
& \forall x_{1} \exists y_{1}Fx_{1}y_{1} \wedge \forall x_{2}\neg Fx_{2}x_{2} & \label{boergerS33b}
\end{eqnarray}
This follows from the fact that $\psi'''$ contains only the following optimized prenex normal form:
\begin{eqnarray}
& \forall x_{1} \exists y_{1}\forall x_{2}(Fx_{1}y_{1} \wedge \neg Fx_{2}x_{2}) &
\end{eqnarray}
$x_{1}$ must be replaced with $y_{1}$ for unification, but C2 of Theorem \ref{b1b2} is satisfied.
\end{exa}

\begin{exa} This example demonstrates how it can be proven that a $\wedge$-NNF is contradictory by deciding upon subformulae $\psi$:
	\begin{eqnarray}
	& \forall x_{1}\exists y_{2}(\neg Gy_{2}y_{2} \wedge \exists y_{1}(\forall x_{3} Fy_{1}x_{1}x_{3} \wedge \forall x_{2}\neg Fx_{2}y_{1}y_{1}) \wedge Gx_{1}y_{2}) & \label{beisp}
	\end{eqnarray}
	Formula (\ref{beisp}) contains the following two pairs of connected literals: $\{Gx_{1}y_{2},\neg Gy_{2}y_{2}\}$ and
	$\{Fy_{1}x_{1}x_{3},$ $\neg Fx_{2}y_{1}y_{1}\}$. $\forall x_{1}\exists y_{2}(\neg Gy_{2}y_{2} \wedge Gx_{1}y_{2})$ is the subformula $\psi$ of the first pair of literals.
    In this case, $\psi=\psi''$, $\psi'''=\{\psi\}$, and $\sigma$ = $\{\{x_{1},y_{2}\}\}$. Hence, C1 is not satisfied. C2, however, is satisfied since $\phi'''$ does not contain any optimized
    prenex normal form with $\forall x_{1}$ in the scope of $\exists y_{2}$. Therefore,
	$\{Gx_{1}y_{2},\neg Gy_{2}y_{2}\}$ is not unifiable. Since both literals do not occur in any other pair of literals from (\ref{beisp}), these literals can be deleted in formula (\ref{beisp}) for efficiency in the search for %proofs by contradiction
	proofs of contradiction. The resulting formula is the subformula $\psi$ of the second pair of literals and is identical to (\ref{embeisp}). The anti-prenex, $\exists$M-optimized formula $\psi''$ is (\ref{embeisp1}). $\sigma$ is unambiguous: $\{\{x_{1},y_{1_{2}}\}, \{x_{2},y_{1_{1}}\}, \{x_{3},y_{1_{2}}\}\}$ (= $\sigma(\ref{embeisp1})$). Therefore, C1 is not satisfied. $\psi'''$ contains only one optimized prenex normal form, which is optimal:
	
	\vspace{0.25cm}
	
	$\begin{array}{lcr}
	& \hspace{1.75cm} \exists y_{1_{2}}\forall x_{1}\exists y_{1_{1}}\forall x_{3}\forall x_{2}(Fy_{1_{1}}x_{1}x_{3} \wedge \neg Fx_{2}y_{1_{2}}y_{1_{2}}) & \hspace{1.15cm} \nonumber (\ref{embeisp3})
	\end{array}$
	
	\vspace{0.25cm}
	
Therefore, C2 is also not satisfied. Applying $\forall$E in accordance with $\sigma(\ref{embeisp1})$ results in the following explicit contradiction:
	\begin{eqnarray}
	& \exists y_{1_{1}}\exists y_{1_{2}}(Fy_{1_{1}}y_{1_{2}}y_{1_{2}}  \wedge \neg Fy_{1_{1}}y_{1_{2}}y_{1_{2}}) &
	\end{eqnarray}
Therefore, the pair of literals $\{Fy_{1}x_{1}x_{3}, \neg Fx_{2}y_{1}y_{1}\}$ in (\ref{beisp}) is unifiable, and consequently, (\ref{beisp}) is contradictory.
\end{exa}

\section{Efficient Identification of Unifiable Pairs of Literals} \label{subkpairs}

The described algorithm for identifying unifiable pairs of literals is cumbersome due to the generation of $\psi$, $\psi'$, $\psi''$ and $\psi'''$.
For the deletion of literals that are not members of any pair of unifiable literals in an NNF $\phi$, it would be desirable to instead apply criteria
that directly refer to $\phi$, such as in the case of identifying pairs of connected literals.
The intention is to specify necessary conditions for unifiable pairs of literals, although these conditions, when taken together, need not be sufficient.
%Für die Entscheidung über Widersprüchlichkeit ist auch gar nicht entscheidend, ob $\cal{D}$ nur noch
%Literale unifizierbar$_{1}$er Literalpaare enthält. Entscheidend ist allein, dass vor Schritt 2 des FOL-Deciders die Menge
%aller unifizierbar$_{1}$er Literalpaare für jedes $\cal{D}$ identifiziert wird. Reduktion der Literale ist also allein
%eine Maxime der Optimierung zum Zwecke höherer Effizienz und Transparenz. Deshalb werden im FOL-Decider in {\tt sat-Reduce} nur k-Paare
%identifiziert, um hierdurch effizient Literale eliminieren zu können, für die ohne viel Aufwand
%entschieden werden kann, dass sie nicht in unifizierbar$_{1}$en Literalpaaren vorkommen. Dies geschieht vor $\exists$M-Optimierung,
%um diese möglichst schlank halten zu können. Erst im Anschluss an $\exists$M-Optimierung wird dann vor Schritt 2 die Menge aller
%unifizierbar$_{1}$er Literalpaare für jedes einzelnen Disjunkt $\cal{D}$ identifiziert.

The following 3 criteria are necessary conditions for unifiable pairs of literals that directly refer to the syntactic properties of a rectified NNF $\phi$.
They can be justified by the fact that the syntactic properties of $\phi$ to which they refer are sufficient for C1 or C2 to be satisfied even after $\exists$M-optimization and the generation of all optimized
prenex normal forms of $\psi'''$.

\begin{description}
	\item[C1U1] This criterion checks whether condition C1 (i.e., an ambiguous substitution list $\sigma$) is satisfied if the $x$ variables in both of a pair of literals are subscripted differently
by using a subscript of -1 in L1 and a subscript of -2 in L2. If the application of Algorithm \ref{sigma} results in an ambiguous substitution list $\sigma$ under this condition, then C1 will be satisfied even after $\exists$M-optimization.
\begin{exa}
This criterion is satisfied if, e.g., an $x$ variable occurs in two positions in one literal and those positions are occupied by different $y$ variables in the other literal; cf. formula (\ref{B1BFormel}).
The criterion is also satisfied if two $x$ variables occur in the same position in both literals and both $x$ variables must be replaced with different $y$ variables due to
their occurrences in other positions; cf. formula (\ref{B1BFormel2}).
\end{exa}
\item[C2U1] This criterion checks whether (i) a universally quantified variable $\nu_{1}$ occurs together with an existentially quantified variable $\mu$ in L1 (resp. L2),
(ii) $\exists \mu$ is in the scope of $\forall \nu_{1}$ in $\phi$ (or $\psi$), (iii) $\nu_{1}$ occupies the same position in L1 (resp. L2) as that of a universally quantified variable
$\nu_{2}$ in L2 (resp. L1), and (iv) $\nu_{2}$ occupies the same position in L2 (resp. L1) as that of $\mu$ in L1 (resp. L2); cf. formula
(\ref{B2AFormel}). In this case, C2 will be satisfied even after $\exists$M-optimization and the generation of all optimized prenexes.
\item[C3U1] This criterion checks whether (i) a universally quantified $x$ variable $\nu_{1}$ must be replaced with an existentially quantified variable $\mu_{2}$,
(ii) a universally quantified $x$ variable $\nu_{2}$ must be replaced with an existentially quantified variable $\mu_{1}$,
(iii) $\nu_{1}$ and $\mu_{1}$ occur in one literal while $\nu_{2}$ and $\mu_{2}$ occur in the other literal, and
(iv) $\exists \mu_{1}$ is in the scope of $\forall \nu_{1}$ while $\exists \mu_{2}$ is in the scope of $\forall \nu_{2}$ in $\phi$ (or $\psi$); cf. formula
	(\ref{B2BFormel}). In this case, C2 will be satisfied even after $\exists$M-optimization and the generation of all optimized prenexes.
\end{description}

Given a set of pairs of connected literals, C1U1-C3U1 enable one to identify a reduced set of pairs of connected literals that is guaranteed to contain all unifiable pairs of literals (= set 1).
However, not all pairs of connected literals within this set are necessarily unifiable.

\vspace{0.3cm}

To identify all and only unifiable pairs of literals, one must generate the anti-prenex, $\exists$M-optimized subformula $\psi''$ for each pair in set 1 and apply the following criteria:

\begin{description}\label{K6U1}
	\item[C4U1] This criterion checks whether the substitution list $\sigma$ of a pair of literals is ambiguous.
	\item[C5U1] This criterion checks whether $\sigma$ contains an $x$ list with an $x$ variable $\nu$ and a $y$ variable $\mu$ such that $\exists \mu$ is in the scope of $\forall \nu$ in $\psi''$.
\end{description}
Given $\psi''$, C4U1 decides whether C1 is satisfied, and C3U1 and C5U1 decide whether C2 is satisfied.\label{T1U1}
The latter is true because the scopes cannot be minimized any further in $\psi''$; cf. Theorem \ref{minmax}.
Given maximally minimized scopes, only two conditions prevent the generation of optimal prenexes:
either $\exists \mu$ is in the scope of $\forall \nu$ in $\psi''$ (case 1), or $\psi''$ is of the form $\ldots (\ldots \forall \nu_{1}\ldots \exists \mu_{1} \ldots A(\mu_{1},\nu_{1}) \wedge
\ldots \forall \nu_{2} \ldots \exists \mu_{2} \ldots B(\mu_{2},\nu_{2}))$ (case 2). Case 1 ensures that $\exists \mu$ will also be in the scope of $\forall \nu$ in any optimized prenex normal form. C5U1 applies to this case. In case 2, any optimized prenex will contain either $\forall \nu_{1}$ to the left of $\exists \mu_{2}$ or $\forall \nu_{2}$ to the left of $\exists \mu_{1}$. Therefore, no optimal prenex exists such that $\exists \nu_{1}$ is to the left of $\forall \nu_{2}$ and $\exists \nu_{2}$ is to the left of $\forall \nu_{1}$. C3U1 applies to this case.
In all other cases, however, either $\forall \nu$ is already
in the scope of the corresponding existential quantifier $\exists \mu$ in $\psi''$, if $\nu$ is to be replaced with $\mu$ for unification, or the strategy applied in Algorithm \ref{wpA} of placing existential quantifiers to the left of universal quantifiers if possible will generate an optimal prenex.
Therefore, it is sufficient to refer to $\psi''$ to decide the unifiability of the pairs in set 1.

C3U1 can already be applied to $\phi$ or $\psi$ independent of the generation of $\psi''$ because $\exists$M-optimization cannot cause C3U1 to not be satisfied in $\psi''$ when it is satisfied in $\phi$ or $\psi$. Hence, it is sufficient to apply criteria C4U1 and C5U1 to the pairs of literals in set 1 to identify the unifiable pairs of literals.

%Dies kann unabhängig und vor der optionalen $\exists$M-Optimierung von $\cal D$ -- die allenfalls erst im letzten Schritt von {\tt start} vorgenommen wird -- geschehen.

Therefore, the following algorithm can be utilized for an optimized proof search by deleting literals in the NNF $\phi$:
\begin{algo}\hfill \label{bereinigung}
	\begin{enumerate}
		\item Generate the set of all pairs of connected literals from $\phi$.
		\item Apply C1U1-C3U1 to this set to minimize it.
		\item For each member of this set, generate the subformula $\phi''$ and apply C4U1 and C5U1 to minimize it. The resulting set A contains all unifiable pairs of literals from $\phi$.
		\item Generate a set B consisting of all literals from $\phi$ that are not contained in set A.
		\item Replace the literals from set B in $\phi$ with the expression `$sat$'. The result is $\phi^{*}$.
		\item Delete all quantifiers in $\phi^{*}$ that do not bind a variable that occurs in some literal. The result is $\phi^{**}$.
		\item Iteratively apply the following rules from right to left, beginning with $\phi^{**}$, as long as the resulting expression is either `$sat$' or no longer contains `$sat$':
		\begin{longtable}{rcll}
			$sat \wedge A$ &  $\dashv\vdash_{sat}$ & $A$ & SAT1\\
			$sat \vee A$ & $\dashv\vdash_{sat}$ & $sat$ & SAT2\\
			\caption{$sat$ Rules}\label{satR}
		\end{longtable}
		The result is $\phi^{***}$: a formula that is sat-equivalent to $\phi$ and contains no literals that cannot contribute to any
		proof of contradiction for $\phi$.
	\end{enumerate}
\end{algo}

\section{Outlook}\label{ausblick}

Given an anti-prenex FOLDNF with disjuncts $D_{i}$ resulting from Algorithm \ref{bereinigung} that (i) are not decidable with regard to their refutability by means of Theorem \ref{herbrandth} because they contain $\vee$ and (ii) cannot be identified as non-contradictory by means of Theorem \ref{dinonwid}, the question arises of how the refutability of such disjuncts $D_{i}$ can be decided. According to the $\wedge$I-minimal proof strategy, this question boils down to the question of how to compute the minimal number of applications of $\wedge$I necessary to unify a minimal set of unifiable pairs of literals in general. Simple examples such as formula (\ref{orkpairbsp}) show that applications of $\wedge$I are indispensable even if one allows for $\exists$M.\footnote{Since applications of $\wedge$I are necessary for
	proofs of contradiction in general, it is convenient to dispense with $\exists$M altogether when one is considering general $\wedge$I-minimal proofs of $D_{i}$ in accordance with the $\wedge$I-minimal proof strategy in the NNF-calculus.}
However, if a disjunct $D_{i}$ is contradictory, then a $\wedge$I-minimal %proof by contradiction
proof of contradiction for $D_{i}$ in the NNF-calculus is guaranteed to exist.
Therefore, it is sufficient to restrict the search for %proofs by contradiction
proofs of contradiction to a systematic search for $\wedge$I-minimal proofs.

According to the $\wedge$I-minimal proof strategy, the necessary applications of $\wedge$I depend on the following parameters:
\begin{enumerate}
\item a minimal set $\tt{L}$ of pairs of unifiable literals  that is sufficient for a proof of contradiction given the unification of its members,
\item a minimal list $\sigma$ of substitutions  that specifies how to replace $x$ variables with $y$ variables to unify the pairs in $\tt{L}$, and
\item optimized prenexes $\wp$ with respect to those quantifiers that bind variables of $\sigma$.
\end{enumerate}
All alternatives for unifying minimal sets of unifiable pairs of literals through the minimal number of substitutions in relation to all kinds of optimized prenexes must be considered in a systematic search for $\wedge$I-minimal proofs. As soon as a minimal set $\tt{L}$ with an unambiguous substitution list $\sigma$ and an optimal prenex $\wp$ is found, a $\wedge$I-minimal proof for $D_{i}$ has been identified.
However, if this does not occur, $\wedge$I must be applied to generate optimal prenexes by stipulating that $x$ variables must be replaced with $y_{0}$ in addition to other substitutions or to separate $x$ variables.

The question is when, in the search for $\wedge$I-minimal proofs, a search path can be abandoned because it is decidable based on purely syntactic criteria that no $\wedge$I-minimal proof can be found along that proof path. A prominent candidate for such a criterion that can be only vaguely specified at this point is as follows: On a certain proof path, as soon as the multiplication of a universally quantified expression through the application of $\wedge$I causes it to be necessary for the universally quantified expression resulting from this multiplication to, in turn, be multiplied by $\wedge$I to achieve similar substitutions of similar unifiable pairs of literals, it can be concluded that such repeated applications of $\wedge$I cannot be a necessary part of a $\wedge$I-minimal proof, and therefore, this proof path can be abandoned. Substitutions and pairs of literals are similar if they are identical in terms of their level-1 indices.
Here, we presume that variables are rectified subsequent to the application of $\wedge$I by increasing the level of their indices by 1; cf. p. \pageref{indexundI}.
The stated conditions can be identified syntactically if one keeps records of all applications of $\wedge$I in which universally quantified expressions are multiplied to unify certain unifiable pairs of literals by means of certain substitutions. I refer to this criterion for abandoning a proof path when applications of $\wedge$I on that proof path induce similar applications of $\wedge$I for similar substitutions in similar unifiable pairs of literals as the `loop criterion'.

For a $\wedge$I-minimal proof strategy, the relevant questions are how to restrict the search for
proofs of contradiction to the search for $\wedge$I-minimal proofs and
at what point the search for such proofs can finally be abandoned on all proof paths in the case of non-contradictory $D_{i}$ due to a criterion such as the loop criterion.
The specific algorithmic definition of a systematic search strategy for $\wedge$I-minimal proofs in fragments of FOL other than $\wedge$I-NNFs will require further detailed investigation and justification.
The intention of this paper was to explain the underlying idea of such a proof search with respect to the most simple example, in which one can dispense with applications of $\wedge$I altogether.
This might encourage, for more general fragments of FOL, a search for decision procedures that are analogous to the presented systematic search, which is based on nothing but purely proof-theoretic reasoning, for $\wedge$I-minimal proofs in the NNF-calculus.


\begin{thebibliography}{N}
%\bibitem[Baaz et al. (2001)]{Baaz}
%Baaz, M., Egly, U., Leitsch, A.: ``Normal Form Transformations'', in: Robinson, A. $\&$ Voronkov, A. (eds.), \emph{Handbook of Automated Reasoning I}, Elsevier, Amsterdam u.a., 2001, pp. 273-333.
\bibitem[Boerger et al. (2001)]{Boerger}
Börger, E., Grädel, E. $\&$ Gurevich, Y.: \emph{The Classical Decision Problem}, Springer 2001.
\bibitem[Fermüller et al. (1993))]{Fermueller1}
Fermüller, C.G., Leitsch, A., Tammet, T., Zamov, N.K.: \emph{Resolution Methods for the Decision Problem. Lecture Notes in Computer Science 679}, Springer 1993.
\bibitem[Fermüller et al. (2001)]{Fermueller2}
Fermüller, C.G., Leitsch, A., Hustadt, U., Tammet, T.: ``Resolution Decision Procedures'', in: Robinson, A. $\&$ Voronkov, A. (eds.), \emph{Handbook of Automated Reasoning II}, Elsevier, Amsterdam u.a., 2001, pp. 1191-1849.
\bibitem[Lampert(2017a)]{Lampert}
	Lampert, T. 2017: ``Minimizing Disjunctive Normal Forms of Pure First-order Logic'', \emph{Logic Journal of the IGPL 25.3}, 325-347.
\bibitem[Lampert(2017b)]{Lampertb}
Lampert, T. 2017: ``Wittgenstein's $ab$-notation: An Iconic Proof Procedure'', \emph{History and Philosophy of Logic 38.3}, 239-262.
\bibitem[Nonnengart $\&$ Weidenbach]{Nonnengart}
		Nonnengart, A. $\&$ Weidenbach, C.: ``Computing small clausal normal forms'', in: Robinson, A. $\&$ Voronkov, A. (eds.), \emph{Handbook of Automated Reasoning I}, Elsevier, Amsterdam u.a., 2001, pp. 335-367.
\bibitem[Quine(1982)]{Quine}
	Quine, W.V.O.: \emph{Methods of Logic}, Fourth Edition, Harvard University Press, Harvard, MA, 1982.
\end{thebibliography}
\end{document}